\documentclass[aps, preprint, nofootinbib,preprintnumbers,eqsecnum,superscriptaddress,sort]{revtex4}
\pdfoutput=1
\usepackage{amssymb,amsmath,amsfonts,amsthm,latexsym,mathrsfs}
\usepackage{graphics, graphicx,color,epsfig,subfigure}
\usepackage[small,hang,justification=raggedright,singlelinecheck=on]{caption}
\usepackage[
      colorlinks=true,
      linkcolor=blue,
      urlcolor=blue,
      filecolor=black,
      citecolor=red,
      pdfstartview=FitV,
      pdftitle={},
        pdfauthor={Gary Horowitz, Nabil Iqbal, Jorge Santos, Benson Way},
        pdfsubject={},
        pdfkeywords={},
        pdfpagemode=None,
        bookmarksopen=true
      ]{hyperref}
\numberwithin{equation}{section}
\def \be {\begin{equation}}
\def \ee {\end{equation}}
\def \bea {\begin{eqnarray}}
\def \eea {\end{eqnarray}}
\def \dd {\mathrm{d}}

\newcommand{\bln}{\begin{align}}
\newcommand{\eln}{\end{align}}
\newcommand{\bst}{\begin{split}}
\newcommand{\est}{\end{split}}
\newcommand{\bi}{\begin{itemize}}
\newcommand{\ei}{\end{itemize}}
\newcommand{\ben}{\begin{enumerate}}
\newcommand{\een}{\end{enumerate}}

\def\le{\left}
\def\ri{\right}
\def\ha{{1\over 2}}

\newcommand{\p}{\partial}

\newcommand\Sig{\Sigma}
\newcommand\lam{\lambda}
\newcommand\Lam{\Lambda}

\def\lam{{\lambda}}

\def\eeq{\end{equation}}

\newcommand\sH{{\ensuremath{{\mathcal H}}}}

\newcommand\sM{{\ensuremath{{\mathcal M}}}}

\newcommand\sV{{\mathcal V}}

\begin{document}
\title {Hovering Black Holes from Charged Defects}

\author{Gary T. Horowitz}
\email{gary@physics.ucsb.edu}
\affiliation{Department of Physics, UCSB, Santa Barbara, CA 93106}

\author{Nabil Iqbal}
\email{n.iqbal@uva.nl}
\affiliation{Institute for Theoretical Physics, University of Amsterdam, Science Park 904, Postbus 94485, 1090 GL Amsterdam, The Netherlands}

\author{Jorge E. Santos}
\email{jss55@cam.ac.uk}
\affiliation{Department of Applied Mathematics and Theoretical Physics, University of Cambridge, Wilberforce Road, Cambridge CB3 0WA, UK \vspace{1 cm}}

\author{Benson Way}
\email{bw356@cam.ac.uk}
\affiliation{Department of Applied Mathematics and Theoretical Physics, University of Cambridge, Wilberforce Road, Cambridge CB3 0WA, UK \vspace{1 cm}}

\begin{abstract}\noindent{
We construct the holographic dual of an electrically charged, localised defect in a conformal field theory at strong coupling, by applying a spatially dependent chemical potential. We find that the IR behaviour of the spacetime depends on the spatial falloff of the potential. Moreover, for sufficiently localized defects with large amplitude, we find that a new gravitational phenomenon occurs: a spherical extremal charged black hole nucleates in the bulk: a hovering black hole. This is a second order quantum phase transition.  We construct this new phase with several profiles for the chemical potential and study its properties. We find an apparently universal behaviour for the entropy of the defect as a function of its amplitude. We comment on the possible field theory implications of our results.
}
\end{abstract}
\maketitle

\tableofcontents

\section{Introduction and Summary}

Over the past several years, gauge/gravity duality has been applied to problems of interest in condensed matter physics, with surprising results. By now, gravitational duals of many condensed matter phenomena have been found, and the gravitational solutions have been used to gain new insight into strongly correlated matter.  

Charged defects are a common feature of many condensed matter systems, with many materials showing great sensitivity to the presence of impurities. Here, we build a gravity dual to an isolated defect at a quantum critical point and study its properties.  (See \cite{Harrison:2011fs,Zeng:2014dra,Erdmenger:2013dpa} for some earlier discussions of a single impurity in a holographic context.)  More precisely, we study a localised electrically charged defect in a strongly coupled $2+1$ dimensional conformal field theory. This is described by a chemical potential 
\be\label{chempot}
\mu(r) = a p(r)\;,
\ee
where we have factored out an overall amplitude $a$ and the profile $p(r)$ vanishes at large $r$. Interesting effects have recently been found in the study of an impurity of this type at linear order about a background with constant $\mu$ \cite{Blake:2014lva}. We will study the nonlinear effects when \eqref{chempot} represents the total chemical potential.

Adding a chemical potential to the CFT corresponds to adding the term $\int d^3x\, \mu(r)\rho(r)$ to the CFT action. The chemical potential has dimension one and the induced charge density $\rho(r)$ has dimension two.  The following simple scaling argument relates the fall-off of $\mu(r)$ to whether this is a relevant, marginal, or irrelevant deformation. If the large $r$ behaviour is $\mu\sim a/r^\beta$, then the dimension of $a$ is $1-\beta$. So one expects that $\beta < 1$ is a relevant deformation, $\beta >1$ is irrelevant,  and $\beta =1$ is marginal. 

The gravitational  dual is a static, axi-symmetric solution of Einstein-Maxwell theory with negative cosmological constant. We focus on solutions at zero temperature.  We construct these solutions perturbatively for small amplitude $a$ and numerically for larger $a$, for several profiles $p(r)$.   We indeed find that the part of the geometry corresponding to the infrared (IR) is determined by the fall-off of the chemical potential.\footnote{The relation between the asymptotic behaviour on the boundary and the near horizon geometry in the bulk has been studied in the case of pure gravity (with no Maxwell field) in \cite{Hickling:2014dra}.} When $\mu(r)$ falls off faster than $1/r$, the zero temperature solution has a standard Poincar\'{e} horizon, as expected for an irrelevant deformation. When $\mu(r) \propto 1/r$ asymptotically,  the $T=0$ solution does not have a standard Poincar\'{e} horizon, but rather an extremal  horizon with nonzero electric flux. This near horizon geometry can be described analytically, and corresponds to a new conformal fixed point in the dual CFT.  When $\mu(r)$ falls-off more slowly than $1/r$, we can find \emph{finite} temperature solutions with a regular black hole horizon in the IR, but the horizon appears to become singular as $T \to 0$. 

The fall-off of the chemical potential also determines the induced total charge. We will see using either field theory or gravitational arguments, that the total charge vanishes when $\mu(r)$ falls off faster than $1/r$, diverges when $\mu(r)$ falls off slower than $1/r$, and is finite and nonzero only when the fall off is $\mu(r)\propto1/r$.

The special marginal case where $\mu(r) = a/r$ everywhere is of particular interest. The corresponding bulk solution can now be found analytically. (It can be obtained by an analytic continuation of a previously known charged, hyperbolic black hole.)  The induced charge density is a delta function at the origin, so we will call this defect the ``point charge".  The exact solution describing the IR geometry of this point charge also describes the IR geometry of all marginal deformations. 

Perhaps our most surprising result concerns what happens when one increases the strength of the defect. In both the irrelevant and marginal cases, as one increases the amplitude $a$, a novel effect takes place: a  spherical extremal charged black hole nucleates in the bulk. The solution remains static and the black hole hovers above the IR horizon, with the electrostatic force towards the boundary balancing the tendency of objects to fall towards the IR horizon.  Near the hovering black hole, the solution looks exactly like that of the standard Reissner-N\"{o}rdstrom-AdS solution. 

The hovering black holes only exist when the amplitude is larger than some critical value $a_\star$, and the size of the black hole goes to zero as $a\to a_\star$ from above. This corresponds to a second order quantum phase transition in the CFT with defect.  For a small range of amplitudes above $a_\star$, solutions exist both with and without black holes, but the black hole solutions dominate in any thermodynamic ensemble. As one continues to increase the amplitude, the size of the hovering black hole continues to grow, without any apparent bound. 

\begin{figure}[th]
\includegraphics[width=.6\textwidth]{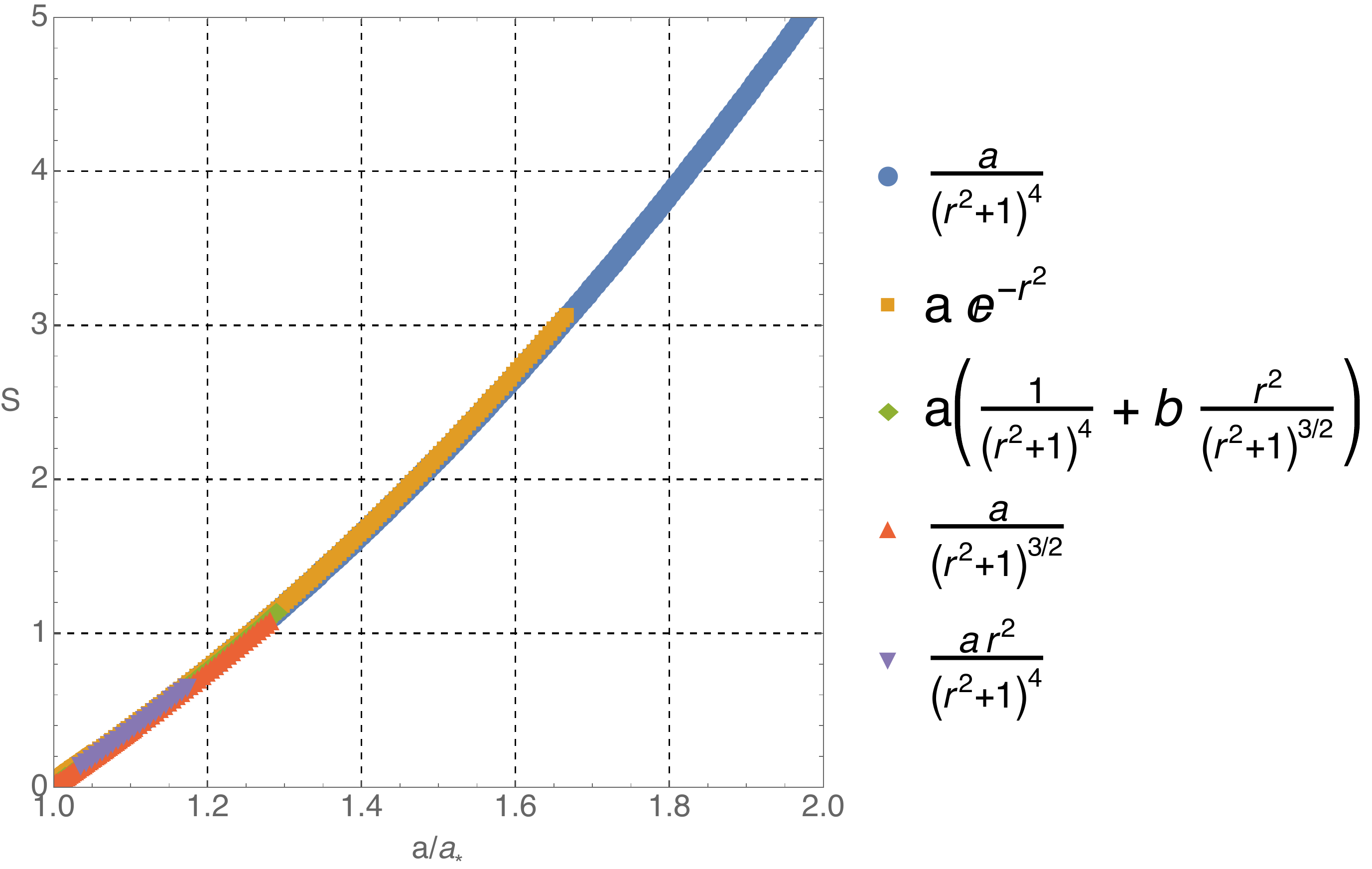}
\caption{\label{fig:entropy1}Entropy of the hovering black hole as a function of $a/a_{\star}$ for several boundary profiles. The different symbols, which are labeled on the right, indicate the various profiles we have considered. One of the profiles has an additional parameter which here is $b=0.075$.  (Here and in the remainder of the paper, we make plots in units of the AdS length $L=1$.)}
\end{figure}

The existence of hovering black holes in the bulk implies that the entropy of the defect increases rapidly with $a$.\footnote{It has been argued that extreme Reissner-N\"{o}rdstrom black holes might be unstable in string theory since they have a large entropy at zero temperature and a diverging density of states \cite{Jensen:2011su}. We will not address this potential complication here.}  In fact, the way this entropy increases with $a$ appears to be universal -- that is, independent of the profile $p(r)$ (provided that it doesn't fall off more slowly than $1/r$). In Fig. \ref{fig:entropy1}, we plot the entropy of the hovering black hole as a function of $a/a_\star$. The different colours represent five different profiles for the chemical potential. The fact that they seem to follow the same curve is remarkable, and not understood. When $a$ is close to $a_\star$ and the black hole is very small, the curve is linear: $S \propto (a-a_\star)$. This is similar to what happens for small black holes in global AdS (but the slope is different). The agreement for larger black holes is mysterious.  We currently have neither a field theory argument nor a gravitational argument that explains this universality.
 
\section{Setup and Closed-form Solutions}

We wish to consider solutions of the Einstein-Maxwell equations that asymptote to $AdS_4$. The action is 
\begin{equation}
S = \frac{1}{16\pi G}\int \mathrm{d}^4 x\,\sqrt{-g}\left[R+\frac{6}{L^2}-F^{ab}F_{ab}\right]\,,
\label{eq:action}
\end{equation}
where $L$ is the AdS length scale and $F\equiv \mathrm{d}A$ is the Maxwell field strength. This action yields the following equations of motion:
\be\label{eq:einsteinmaxwell}
G_{ab}\equiv R_{ab}+\frac{3}{L^2}g_{ab} - 2\left(F_{ac}F_{b\phantom{c}}^{\phantom{b}c}-\frac{1}{4}g_{ab}F^{cd}F_{cd}\right)=0\;,\qquad
\nabla_a F^{ab}=0\;. 
\ee

We are interested in static, axisymmetric solutions.  Therefore, there is a timelike Killing vector $\partial_t$ and an axisymmetric Killing vector $\partial_\phi$.  Our solutions will depend upon the remaining two coordinates (i.e., the problem is cohomogeneity two).  The field theory metric is conformal to the metric on the AdS boundary.  We choose the boundary to be conformal to Minkowski space
\be\label{eq:bndisflat}
\dd s^2_\partial = -\dd t^2+\dd r^2+r^2 \dd\phi^2\;.
\ee
The chemical potential on the field theory is given by the gauge field on the boundary.  We choose some axisymmetric profile for the gauge field
\be\label{eq:bndgaugefield}
A|_\partial =\mu(r)\mathrm{d}t\;,\qquad \lim_{r\to\infty}\mu(r)=0\;,
\ee
where the second condition is placed to model a localised defect.  We shall see that many physical properties of our solutions will depend upon the falloff.

We are therefore searching for regular solutions to \eqref{eq:einsteinmaxwell} satisfying the conditions \eqref{eq:bndisflat} and \eqref{eq:bndgaugefield}.  The solution with $\mu=0$ (i.e. the vacuum solution) with the above conditions is of course $AdS_4$ in Poincar\'e coordinates
\be\label{eq:poincare}
\dd s^2_{AdS}=\frac{L^2}{z^2}\left(-\dd t^2+\dd r^2+r^2\dd\phi^2+\dd z^2\right)\;,\qquad A=0\;.
\ee
As it turns out, there is also an analytic solution for (at least) one other profile for $\mu(r)$ which we will discuss shortly.  If we relax the second condition in \eqref{eq:bndgaugefield}, there is also the well-known Reissner-N\"{o}rdstrom-AdS solution for a constant boundary profile $\mu(r) \equiv \mu_0$.

\subsection{Point charge conformal defect}\label{sec:pointcharge}
Consider a boundary profile with the chemical potential
\be\label{eq:pointchargeprofile}
\mu(r) = \frac{a}{r} \ .
\ee
For this (and only this) choice of falloff the parameter $a$ is dimensionless, and thus there are no scales in this problem.  In fact, this choice of boundary chemical potential breaks boundary translations but preserves an $SO(2,1) \times SO(2)$ subgroup of the full $SO(3,2)$ symmetry group of the conformal vacuum \eqref{eq:poincare}. Importantly, there is a preserved scaling symmetry, which scales time while simultaneously scaling towards $r=0$ on the boundary. 

Note that the chemical potential is singular at the origin, which (as we will see) can be interpreted as the presence of a {\it conformal defect} at that point. This conformal defect should be viewed as an IR fixed point that can govern the low-energy physics obtained when a translation-breaking chemical potential is applied. The properties of this defect are universal data characterising the CFT and are calculable. A similar electric defect was studied in the $O(N)$ model in \cite{metlitski2008valence}: interestingly, our results obtained from gravity are qualitatively similar to those obtained therein. In \cite{Dias:2013bwa} a very similar fixed point was also argued to govern the IR physics of a vortex in a holographic superconductor.

\begin{figure}[th]
\begin{center}
\includegraphics[width=.4\textwidth]{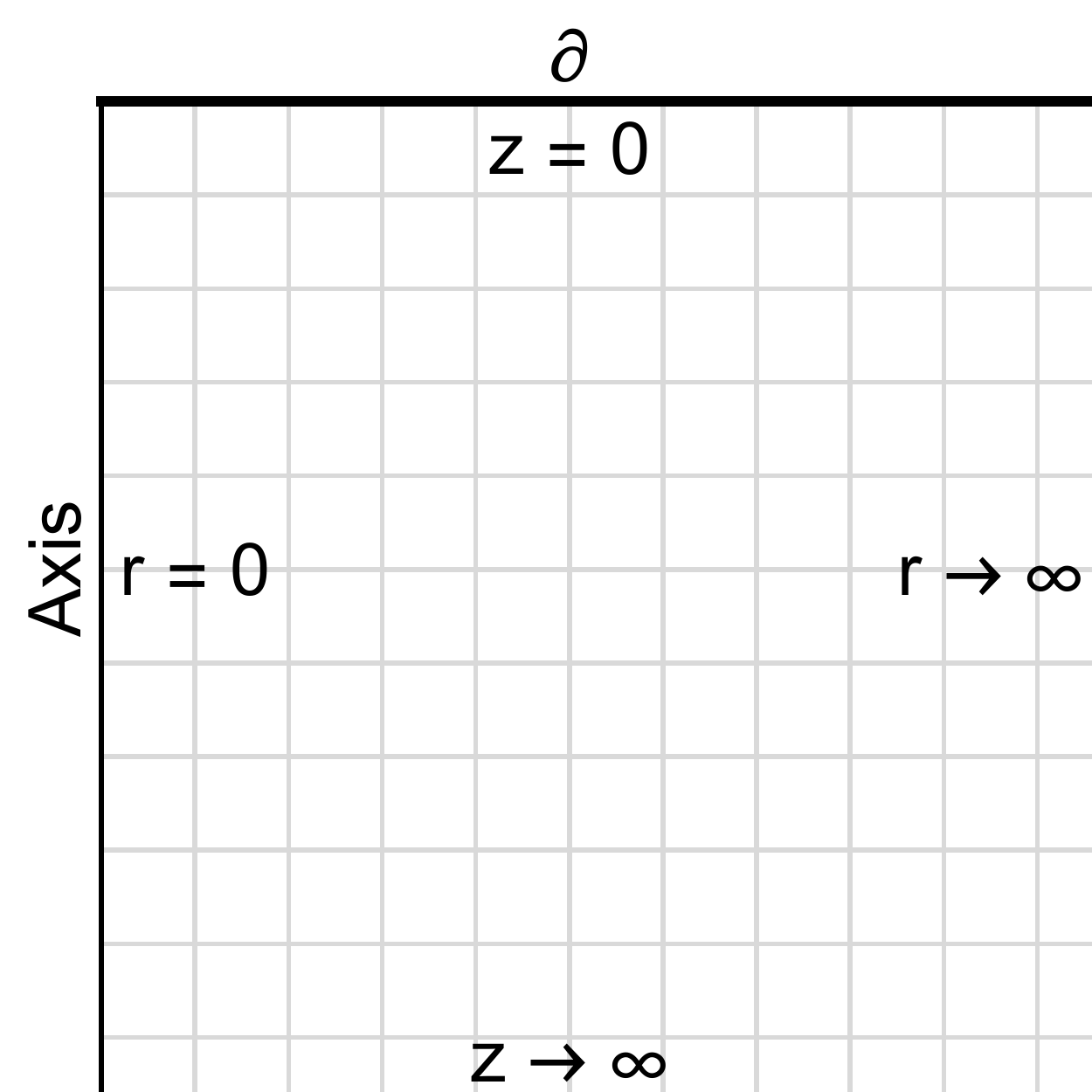}\hspace{1cm}
\includegraphics[width=.4\textwidth]{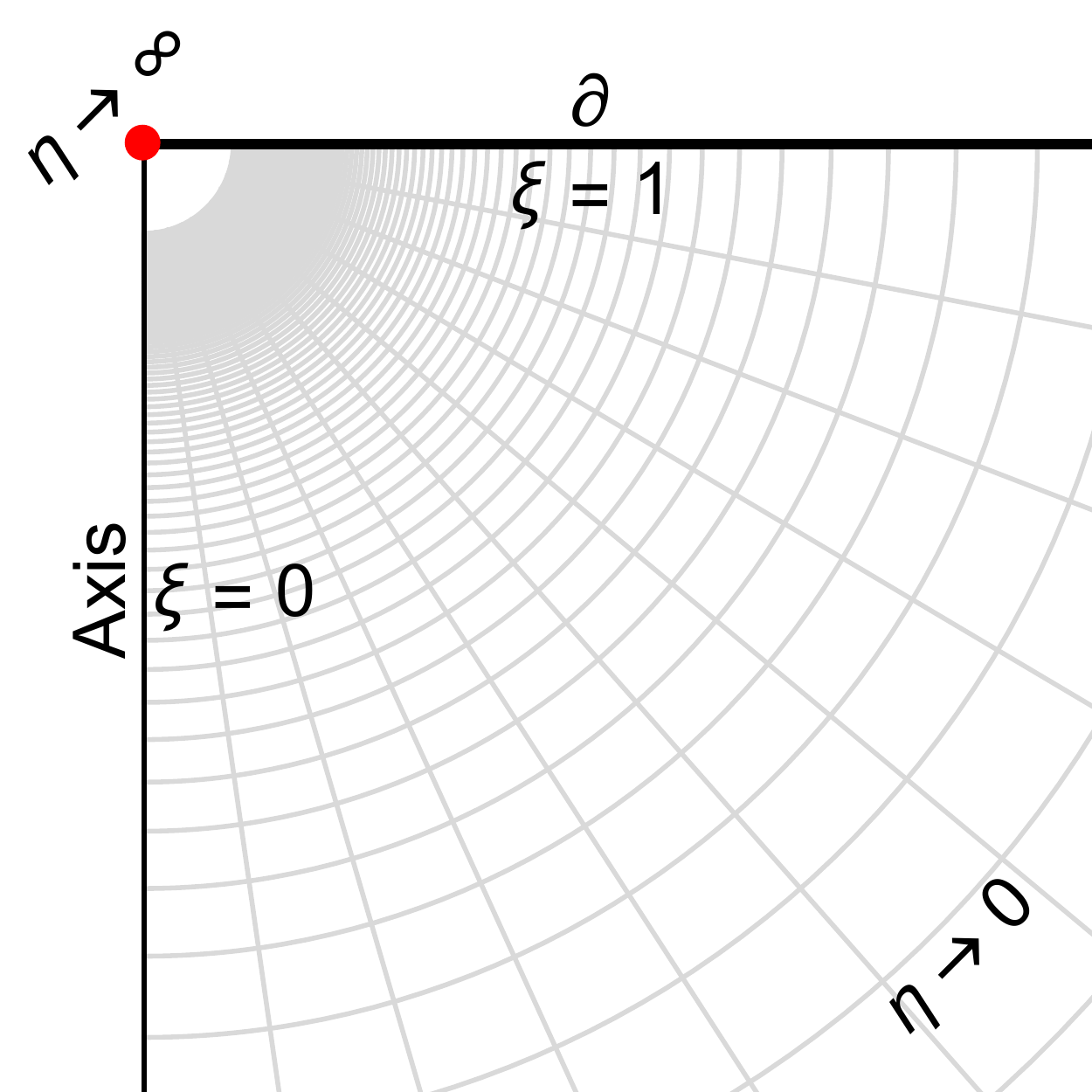}
\end{center}
\caption{Sketches for the two coordinate systems (left) Eq.~(\ref{eq:poincare}) and (right) Eq.~(\ref{eq:polar}).}\label{fig:1}
\end{figure}

We would like to describe the bulk geometry corresponding to \eqref{eq:pointchargeprofile}. But first, it would be convenient to work in coordinates that make the preserved symmetry manifest.  First, note that $\mathbb{R}^{2,1}$ is conformal to $AdS_2 \times S^1$:
\be
-\mathrm{d}t^2 + \mathrm{d}r^2 + r^2 \mathrm{d}\phi^2 = r^2\le(\frac{-\mathrm{d}t^2 + \mathrm{d}r^2}{r^2} + \mathrm{d}\phi^2\ri)\;.
\ee
The preserved subgroup $SO(2,1) \times SO(2)$ is now realised geometrically by the isometry group of $AdS_2 \times S^1$. The defect at the origin is located at the boundary of $AdS_2$. We would now like to write $AdS_4$ so that it is foliated by such a slicing rather than by $\mathbb{R}^{2,1}$. Notice that the $r$ and $z$ components in \eqref{eq:poincare} form a conformally flat subspace written in Cartesian coordinates.  We simply perform a Cartesian to polar transformation, introducing a new radial coordinate $1/\eta$ and angular coordinate $\xi$ related to the usual polar angle by $\cos\theta = 1-\xi^2$:
\be\label{eq:transf}
r = \frac{\xi\sqrt{2-\xi^2}}{\eta}\;,\qquad z= \frac{1-\xi^2}{\eta}\,,
\ee
which gives us the following line element for pure $AdS_4$
\be\label{eq:polar}
 \mathrm{d}s^2_{AdS} = \frac{L^2}{(1-\xi^2)^2}\left[-\eta^2\mathrm{d}t^2+\frac{\mathrm{d}\eta^2}{\eta^2}+\frac{4\mathrm{d}\xi^2}{2-\xi^2}+\xi^2(2-\xi^2)\mathrm{d}\phi^2\right]\,.
\ee
Here $(t,\eta)$ form the $AdS_2$ factor. The conformal boundary is located at $\xi=1$, and the boundary metric is now $AdS_2 \times S^1$, as desired. The origin of the boundary, $r=z=0$, has been mapped to $\eta\rightarrow\infty$, which is itself the timelike boundary of this new $AdS_2$ factor.  The symmetry axis at $r=0$ is now located at $\xi=0$, and the $\phi$ circle smoothly closes off there with periodicity $2\pi$. The Poincar\'e horizon ($z\rightarrow\infty$) is now at $\eta=0$ and by construction now connects to the boundary.  See Fig.~\ref{fig:1} for a pictorial representation of this coordinate system. 

Now by turning on an appropriate gauge field, we can find an exact charged solution to the Maxwell-Einstein system \eqref{eq:einsteinmaxwell}. We will call this solution the \emph{point charge}, for reasons that will become clear. Since we want to keep the symmetries of the $AdS_2 \times S^1$, the solution will only depend nontrivially on $\xi$. The line element and gauge field are given by
\be
 \mathrm{d}s^2 = \frac{L^2}{\lambda^2(1-\xi^2)^2}\left[-\eta^2\mathrm{d}t^2+\frac{\mathrm{d}\eta^2}{\eta^2}+\frac{4\lambda^2\mathrm{d}\xi^2}{f(\xi)}+\xi^2 f(\xi)\mathrm{d}\phi^2\right]\;,\qquad A=L \,a_\lambda\eta\mathrm\,\mathrm{d}t\,, \label{pcharg}
\ee
where $1\leq\lambda\lesssim4.43$ is a constant, and
\be\label{deffa}
f(\xi)=2-\xi^2+(\lambda-1)(1-\xi^2)^2(2-(\lambda+3)\xi^2)\;,\qquad a_\lambda=\frac{\sqrt{(\lambda-1)(\lambda+3)}}{\lambda^2}\;.
\ee
This solution can be obtained by first performing a double Wick-rotation of a magnetically charged hyperbolic black hole \cite{Cai:2004pz} in $AdS_4$.  The parameters can then be tuned to remove singularities, and we are left with the above one-parameter family of solutions.  This solution has appeared earlier in the literature (in different coordinates) and shown to be essentially the unique near horizon geometry for a smooth extremal horizon in AdS \cite{Kunduri:2013gce}. We are now interpreting it as the entire bulk geometry.

The gauge field has no dependence on the holographic direction $\xi$, and corresponds  to a constant electric field of magnitude $a_{\lam}$ pointing along the radial direction of the $AdS_2$. In the limit $\lam \to 1$, the gauge field vanishes and this solution approaches vacuum $AdS_4$ as written in \eqref{eq:polar}. Note from \eqref{deffa} that there are two values of $\lambda$ which give the same $a_\lambda$, meaning that there are two branches of solutions, as well as a maximum value at $a=a_{\lam\mathrm{max}}$. We will comment on the relative interpretation of these two branches shortly. 

What is the charge of this solution? Via the normal rules of AdS/CFT, the field theory current $j^{\mu}$ is defined in terms of a functional derivative\footnote{Rather than use the normalisation of the action given in \eqref{eq:action}, here we have simply picked a convenient normalisation for the current to minimise factors in later formulas.},
\be
\langle j^{\mu}(x) \rangle = -\frac{\delta S}{\delta a_{\mu}(x)} = \frac{1}{4\pi} \sqrt{-g_{\p\sM}}F^{a\mu}N_a\;,
\ee
where $a_{\mu}$ is the boundary value of the bulk gauge field, $N_a$ is a normal vector to the boundary, and $g_{\p\sM}$ refers to the metric on the boundary, including all conformal factors. This means that the total charge can be written as
\be
Q = \int_{\Sig} d^{2}x\;n_{\mu} \frac{\delta S}{\delta a_{\mu}(x)}\;,
\ee
with $\Sig$ a spacelike slice on the boundary and $n_{\mu}$ a timelike vector normal to this spacelike slice. 

\begin{figure}[th]
\begin{center}
\includegraphics[width=.4\textwidth]{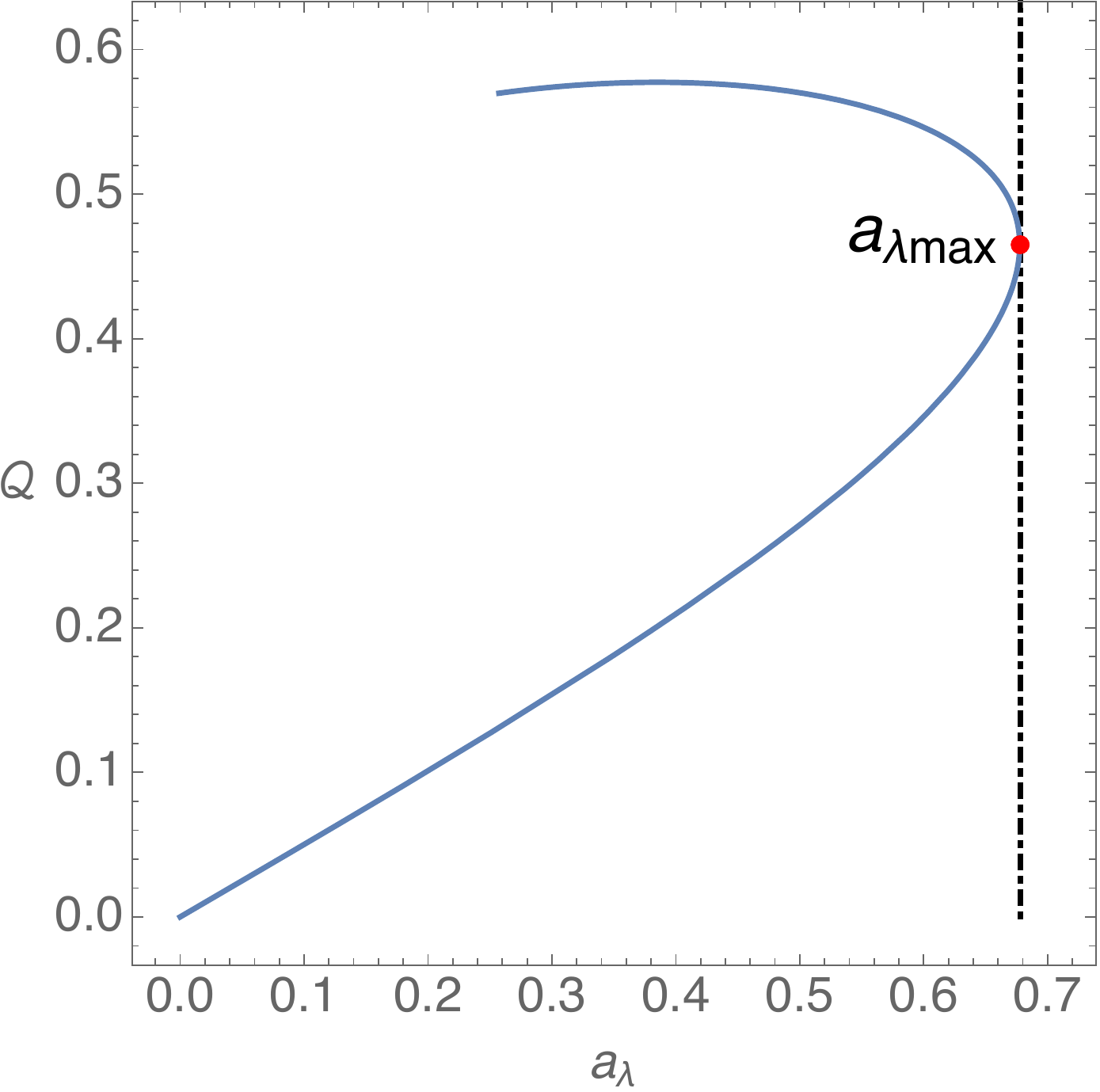}
\end{center}
\caption{Charge $Q$ on defect as a function of applied electric field $a_{\lam}$. The lower branch of solutions is continuously connected to vacuum $AdS_4$ with zero charge and applied field.}\label{fig:Qpointcharge}
\end{figure}

In most circumstances the boundary at infinity has only one component. In our case, however, it actually has two: we have the usual conformal boundary as $\xi \to 1$, but we also have the boundary of the $AdS_2$ factor at $\eta \to \infty$, extending along all $\xi$. Only the latter component contributes to the charge, which we can now explicitly evaluate to be
\be
Q = \frac{1}{4\pi}\int_{\p {AdS_2} \times S^1} d\xi d\phi\, \sqrt{g_{\Sig}}\sqrt{-g_{tt}g_{\eta\eta}}F^{t\eta} = \frac{1}{2}\lam L a_{\lam}\;.
\ee
A plot of the charge as a function of the applied electric field is shown in Fig.~\ref{fig:Qpointcharge}. 

In the boundary theory, this charge is localised at the boundary of the $AdS_2$. It is helpful to interpret this in the original $\mathbb{R}^{2,1}$ conformal frame: if we map back to the coordinates \eqref{eq:bndisflat}, the chemical potential on the boundary is indeed $\mu(r)=a_\lambda/r$, and the charge computed above arises from a delta function contribution to the field theory charge density localised at the origin: $\langle \rho(x) \rangle = Q \delta^{(2)}(x)$. Thus we have a finite charge bound to the defect. It is because of this delta function that we refer to this solution as the ``point charge".

Next, we turn to the entropy. The existence of the $AdS_2$ endows the bulk solution with an extremal horizon at $\eta = 0$, which extends from $\xi = 0$ to the boundary at $\xi = 1$, and whose associated entropy is
\be
S(\lam) = \frac{1}{4G_N} \int_{\sH} d\xi d\phi \sqrt{g_{\xi\xi}g_{\phi\phi}} = \frac{\pi L^2}{G_N \lam} \int_0^{\xi_{\Lam}} d\xi \frac{\xi}{(1-\xi^2)^2}\;, \label{Sexp}
\ee
where we have cut off the $\xi$ integral at a UV cutoff $\xi_{\Lam} \sim 1$. As this horizon intersects the boundary in a circle that surrounds the defect, it should actually be interpreted as an {\it entanglement entropy} computed via the usual Ryu-Takayanagi prescription \cite{Ryu:2006bv}: indeed every constant-$\eta$ slice, including that at $\eta \to \infty$, is a bulk minimal surface. 

\begin{figure}[th]
\begin{center}
\includegraphics[width=.5\textwidth]{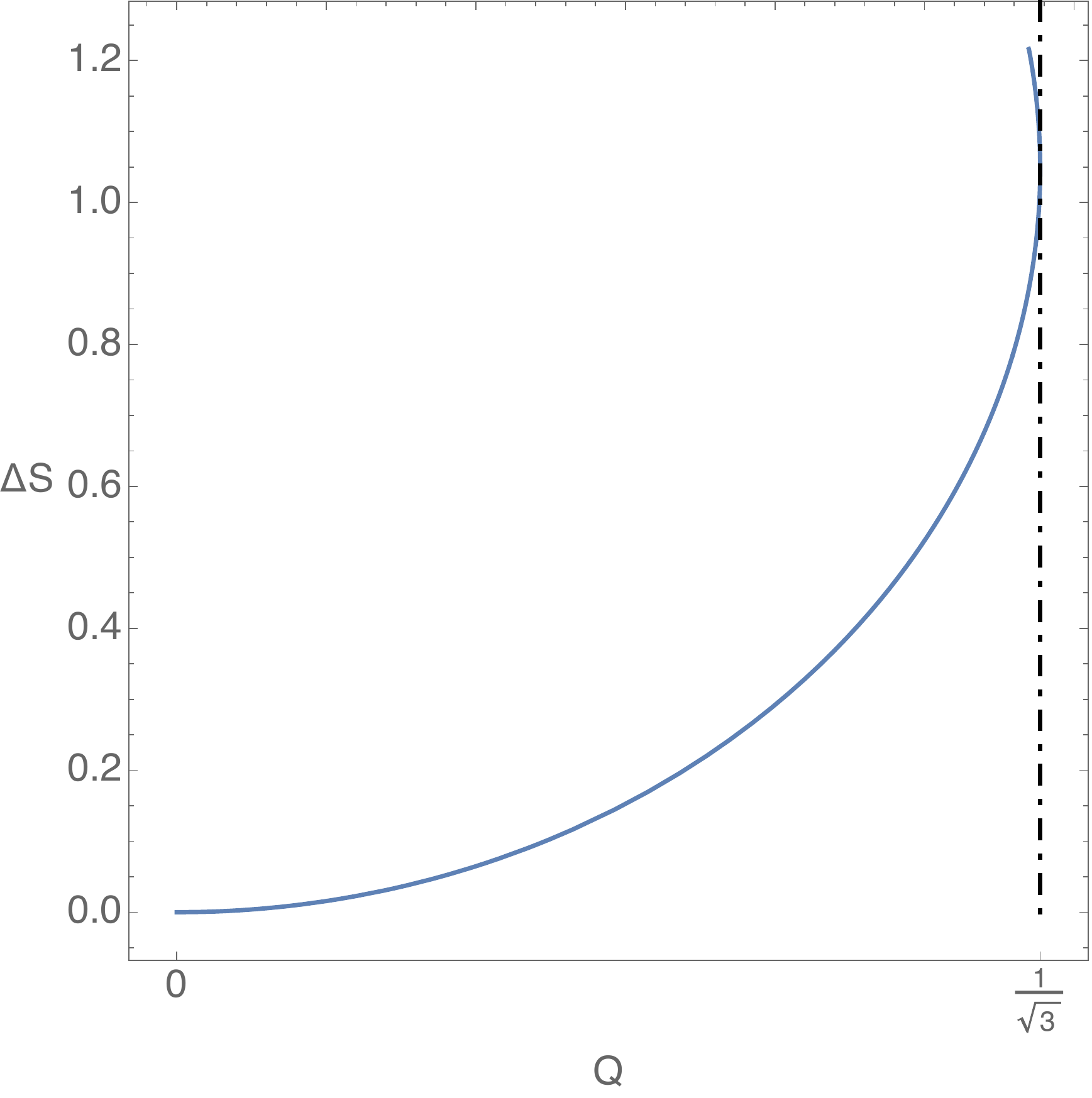}
\end{center}
\caption{Regulated defect entropy for the two branches of the point-charge solution as a function of defect charge $Q$. Note the existence of a maximum charge.}\label{fig:Spointcharge}
\end{figure}

Thus, we are computing the entanglement entropy of the defect with its surroundings. This {\it defect entropy} \cite{PhysRevLett.67.161} is a well-studied object in two dimensions; see e.g. \cite{Jensen:2013lxa} for a discussion of the higher dimensional case. The UV divergence in \eqref{Sexp} is thus the usual UV divergence of the entanglement entropy: we can obtain a finite answer by subtracting the same entanglement entropy computed without the defect present, i.e. with $\lam \to 1$. As usual, some care must be taken in the matching of cutoffs in this subtraction. By ensuring that the $\phi$ circle has the same asymptotic size at the cutoff, we obtain for the regulated entropy
\be
\Delta S(\lam) = S(\lam) - S(\lam = 1) = \frac{\pi L^2}{2 G_N}\le(1 - \frac{1}{\lam}\ri)\;.
\ee
A plot of the regulated entropy versus the total charge is displayed in Fig.~\ref{fig:Spointcharge}. 

Finally, we note that while the solutions are uniquely labeled by $\lam$, there are two values of $\lam$ that give the same $a_{\lam}$, meaning that there are two branches of solutions that meet at the maximum value of $a_{\lam\mathrm{max}} \approx 0.678$ at $\lam_{a\mathrm{max}} = \ha(\sqrt{33} - 3) \approx 1.37$. Clearly only one of these branches (the one with $\lam < \lam_{a\mathrm{max}}$) is continuously connected to vacuum $AdS_4$ at $\lam = 1$, as shown in Fig.~\ref{fig:Qpointcharge}. Since this branch has a smaller charge, we will call it the ``lower branch'', and the other the ``upper branch''. One can show that for $\lam > \lam_0 \approx 4.43$, $f(\xi)$ develops extra zeros in the domain of interest, so $\lam \in [1,\lam_0)$. 

We have performed a preliminary investigation of the perturbation spectrum around this solution. These fluctuations can be classified by their conformal dimensions under $AdS_2$ scaling. As it turns out, the upper branch with $\lam > \lam_{a\mathrm{max}}$ supports a perturbation in the scalar channel that is {\it relevant} with respect to $AdS_2$ scaling, meaning that it is unstable in the RG sense. This operator becomes marginal precisely at $\lam = \lam_{a\mathrm{max}}$, where its existence can be understood in terms of an infinitesimal variation of the charge without changing the boundary conditions on $a_{\lam}$, and is irrelevant in the lower branch for $\lam <\lam_{a\mathrm{max}}$. Thus in the absence of fine-tuning in the UV, we expect only the lower branch to be realised in physical applications, as we will explicitly find in the remainder of this paper. 
  
\section{Expectations}

Before undertaking a detailed gravitational analysis for more general chemical potentials, we first discuss some expectations for the results based on simple analytical arguments. 

\subsection{Relevance of chemical potential and the total charge}\label{sec:falloffs}
Recall that the central problem of this paper is to study a boundary profile for the chemical potential of the form
\be
\mu(r) = a p(r) \label{chempotrpt}\;,
\ee
with $p(r)$ some choice of profile function. Normally a chemical potential is always a relevant deformation, but this is the case only for a homogenous potential. Let us assume that at large $r$, $p(r)$ behaves as a power law, $p(r) \sim r^{-\beta}$. Then, as mentioned in the introduction, the dimension of $a$ is $1-\beta$: we can conclude immediately that for $\beta > 1$ this constitutes an {\it irrelevant} perturbation, but for $\beta < 1$ the perturbation is {\it relevant} and we should flow to a new fixed point. For the precise value $\beta = 1$, $a$ is dimensionless and the perturbation appears marginal. In this case we can actually construct a line of IR fixed points (parametrized by $a$) explicitly in gravity: these are the point charge solutions described above. 

We would now like to understand how we expect the induced charge density $\langle \rho(r) \rangle$ to behave in the presence of such a chemical potential. The full dependence on $r$ will clearly depend on details; in this section we will attempt only to determine its asymptotic falloff with $r$. At large $r$ the chemical potential is small, and one might expect a linear-response analysis about the vacuum to be valid, where schematically the charge density obeys $\langle\rho(x)\rangle \sim \int d^3y \langle \rho(x) \rho(y) \rangle \mu(y)$. In the Euclidean vacuum we have $\langle \rho(x)\rho(y)\rangle \sim k |x-y|^{-4}$, where $k$ is a constant that counts the number of charged degrees of freedom in the CFT. 

Thus in the presence of a chemical potential that is static, we can perform the integral over Euclidean time to find that the charge density should behave as
\be
\langle \rho(\vec{x}) \rangle \sim \int d^2y \frac{k \mu(\vec{y})}{|\vec{x}-\vec{y}|^3} \approx k\int d^2y \mu(y) \frac{1}{|\vec{x}|^3} \label{rhodist}\;,
\ee
where all arguments are now purely spatial, and in the last equality we have assumed $|\vec{x}| \gg |\vec{y}|$, i.e. we are far outside the core\footnote{Note that naively there is a UV divergence in this expression arising from the short-distance behavior of the correlator when $x$ approaches $y$. There is actually a delta function contact term present in the vacuum correlator. The coefficients of such contact terms are often thought to be scheme-dependent: however in this case the requirement that the total charge operator annihilates the vaccum fixes the coefficient of this delta function to precisely cancel the UV divergence in (3.2).}. The behavior of this integral depends on the value of $\beta$. 

For $\beta > 2$, the integral is divergent in the UV.  This means that the integral will be cut off by a length scale $R_\Lam$ coming from the structure of the profile function $p(r)$ at small $r$, and we find
\be
\langle \rho(r)\rangle \sim k\frac{a R_\Lam^{2-\beta}}{r^3} \qquad \beta > 2 \ . \label{rhofast}
\ee
In particular, note that an arbitrarily well-localized charge distribution still sources a power-law tail in the induced charge $r$, and even communicates details about the core of the distribution (stored in the existence of the scale $R_\Lam$) to arbitrarily long distances. This is due to the long-range correlations present in the vacuum of the CFT. 

For $\beta < 2$, the integral instead appears to diverge in the IR: however here we are using the wrong integrand, as the second approximation $|\vec{x}| \gg |\vec{y}|$ in \eqref{rhodist} is invalid. We should instead cut off the integral at $|\vec{y}| \sim |\vec{x}|$ to find
\be
\langle \rho(r) \rangle \sim k\frac{a}{r^{\beta+1}} \sim \frac{k \mu(r)}{r} \qquad \beta < 2\;. \label{rhoslow}
\ee
For $\beta = 2$, we expect to find extra logarithmic factors. 

We now use these results to demonstrate an interesting fact: in a CFT, the net induced charge by a sufficiently localized chemical potential is {\it zero}. The basic idea is that current is conserved, and thus to accumulate a charge in the interior we must pull charge from infinity, so the chemical potential must fall off sufficiently slowly to make this possible. Consider starting in the vacuum with $a = 0$ in \eqref{chempotrpt} and slowly increasing the chemical potential by making $a(t)$ a slowly varying function of time. In the case $\beta < 2$, note that current conservation $\p_{\mu} j^{\mu} = 0$ together with \eqref{rhoslow} tells us that at large $r$ the radial inflow of current satisfies
\be
D_r \langle j^r \rangle = - \p_t \langle \rho \rangle \sim k\frac{\dot{a}(t)}{r^{\beta+1}} \ . 
\ee
Now the net charge accumulated in the interior is equal to the total flux of current through a large circle at infinity, i.e.
\be
\frac{dQ}{dt} = \lim_{r \to \infty} r \oint d\phi j^r \sim \lim_{r \to \infty} k\frac{\dot{a}(t)}{r^{\beta-1}} \ . 
\ee
So if we also have $\beta > 1$, then this flux is zero. This conclusion also holds for the faster falloff with $\beta>2$ \eqref{rhofast}. Thus any irrelevant chemical potential cannot pull charge from infinity and will only redistribute the charge that is already present in the vacuum, meaning that the {\it net} charge will always vanish. On the other hand, for a {\it relevant} chemical potential with $\beta < 1$, we find from \eqref{rhoslow} that the total charge diverges. Thus, the only way to obtain a finite and nonzero amount of charge is with a precisely marginal profile, which is the case for the point charge solution studied above. 

We also note that simple generalisations of the arguments above also allow us to predict the asymptotic falloff of other quantities, e.g. the energy density $\langle T^{t}_{\phantom{t}t} \rangle$. To determine this following the logic leading to \eqref{rhodist} we now need to consider a three-point function, as the two-point function $\langle T \rho \rangle$ vanishes. We thus have an expression of the schematic form $\langle T(x) \rangle \sim \int d^3 y_1 d^3 y_2 \langle T(x) \rho(y_1) \rho(y_2) \rangle \mu(y_1) \mu(y_2)$. The precise form of the three-point function is complicated, but we know that its total mass dimension is $7$.  Performing the integrals over $t_1$ and $t_2$ above will reduce this dimension down to $5$. If we can now assume in the integral above that $|\vec{x}| \gg |\vec{y}_1|, |\vec{y}_2|$, then we find the analog of \eqref{rhodist} for the energy density to be:
\be
\langle T^{t}_{\phantom{t}t}(x) \rangle \sim C \int d^2 y_1 d^2 y_2 \frac{1}{|\vec{x}|^5} \mu(y_1) \mu(y_2) , 
\ee
where $C$ is a constant. The integrals above converge in the IR if $\beta > 2$, and so we find
\be\label{energyfalloff}
\langle T^{t}_{\phantom{t}t}(r) \rangle \sim \frac{C}{r^5} \qquad \beta > 2\;,
\ee
whereas if $\beta < 2$ the integrals over $y$ are IR divergent and should be cut off where we are evaluating the energy, leading to
\be
\langle T^t_{\phantom{t}t}(r) \rangle \sim \frac{C}{r^{2\beta + 1}} \qquad \beta < 2 \ .
\ee
The above reasoning is precisely the same as for the charge density. We note that the above expressions may receive extra logarithmic factors in $r$, as the dimension of the current and energy are both integers; indeed through explicit perturbative calculations we do find such logarithmic corrections when $\beta > 2$. 

\subsection{Charged geodesics and the existence of hovering black holes}\label{sec:geodesics}

For generic $\mu(r)$, the bulk Einstein-Maxwell solution is not known analytically, and we will find it numerically. Given such a solution
  without a hovering black hole, how could one determine if a small spherical black hole can be added and remain static?  Sufficiently small extremal black holes behave essentially like test particles in a background.  Therefore, we can search for static time-like orbits for charged particles. These correspond to stationary points of the geodesic equation coupled to a Lorentz force:
\begin{equation}
U^a\nabla_a U_b = \frac{q}{m}F_{ab}U^b\quad\text{with}\quad U^aU_a = -1\,,
\label{eq:fixed}
\end{equation}
where $q$ is the particle charge and $m$ its mass. A similarly motivated study of probe orbits was performed in a gauged supergravity model in \cite{Anninos:2013mfa}. For static spacetimes, such as the ones we are considering, we can readily integrate this equation. The fixed points of Eq.~(\ref{eq:fixed}) will correspond to local extrema of the following potential
\begin{equation}
\mathcal{V} = \sqrt{-g_{tt}}-\frac{q}{m}A_t\,.
\label{eq:potgeo}
\end{equation}
Note that the normalisation of $F_{ab}$ in the action (\ref{eq:action}) was chosen so that $\mathcal{V}$ is identically zero for extremal ($|q| = m$) particles in flat space.  

By symmetry, it is easy to show that any minima or maxima of $\mathcal V$ must lie on the axis of symmetry.  It is also clear from the form of $\mathcal{V}$  that the particles for which the orbits will form first must maximise $|q|/m$. Since Reissner-N\"{o}rdstrom (RN) black holes in flat space must have $|q|\leq m$, we focus on the extremal case, for which $|q| = m$.  Our task is therefore to study extrema of (\ref{eq:potgeo}) for extremal particles along the axis as a function of the holographic direction. 

At a minimum of the potential, one can expect to place a small (extremal) particle.  Yet, we now argue that \emph{static} hovering black hole solutions exist only if this potential $\mathcal V$ has a minimum below zero (not necessarily just when a minimum exists).  Essentially, small static black holes behave as though they are in flat space, and $\mathcal V$ must be zero for extremal particles in flat space.  As an instructive example, let us compute this potential for \emph{global} AdS and ask when an extremal RN black hole forms.  In this case, the chemical potential is constant and equal to $\mu$. Our potential would reduce to:
\begin{equation}
\mathcal{V}_{\mathrm{global\;AdS}}(r) = \sqrt{1+\frac{r^2}{L^2}}-\mu\,.
\end{equation}
Clearly, there is a minimum in the potential for any value of $\mu$. Yet, the entropy of extremal RN black holes in global AdS is given by
\begin{equation}
\mathcal S  = \frac{\pi\,L^2}{3}(\mu^2-1)\,,
\label{eq:entropy}
\end{equation}
So small black holes have $\mu \approx 1$, where the minimum of $\mathcal{V}_\mathrm{global\;AdS}$ crosses zero. 

Finally, let us compute the potential for the point charge solution \eqref{pcharg}.  It is given by
\be
\mathcal V_{\mathrm{point}}\propto\frac{\lambda-\sqrt{(\lambda-1)(\lambda+3)}}{\lambda^2}\eta\;.
\ee
The potential therefore has constant slope and does not develop extrema.  

\section{Numerical Construction}

In this section, we describe our numerical construction of these solutions.  The reader who is uninterested in numerical details may freely skip this section.  We opt to use the DeTurck method, first introduced in \cite{Headrick:2009pv} and studied in great detail in \cite{Figueras:2011va}. The method first requires a choice of reference metric $\bar g$ that is compatible with the boundary conditions. One then solves the Einstein-Maxwell-DeTurck equations
\begin{equation}\label{eq:EMD}
G^{H}_{ab}\equiv G_{ab}-\nabla_{(a}\xi_{b)}=0\;,\qquad \nabla_a F^{ab}=0\;,
\end{equation}
where $\xi^\mu=g^{\rho\sigma}\left[\Gamma^{\mu}_{\rho\sigma}(g)-\Gamma^{\mu}_{\rho\sigma}(\bar g)\right]$, and $\Gamma^{\mu}_{\rho\sigma}(\mathfrak g)$ is the Levi-Civita connection for a metric $\mathfrak g$.  

These equations are identical to the Einstein-Maxwell equations (\ref{eq:einsteinmaxwell}) with an additional DeTurck term $\nabla_{(a}\xi_{b)}$.  The new term produces non-degenerate kinetic terms for all metric components and automatically fixes the gauge $\triangle x^\mu = g^{\rho\sigma}\Gamma^\mu_{\rho\sigma}(\bar g)$, a generalisation of Harmonic gauge.  In addition, for the systems considered in this paper, one can show that the Einstein-Maxwell-DeTurck equations are an elliptic system of PDEs \cite{Headrick:2009pv}.  

It is clear from inspection that any solution to $G_{ab}=0$ with $\xi =0$ is a solution of $G^H_{ab}=0$. The converse, however, is not necessarily true. For certain types of problems, it is possible to prove that solutions with $\xi\neq0$, coined DeTurck solitons, cannot exist \cite{Figueras:2011va}. For the case at hand, we do not have such a proof; the proof in \cite{Figueras:2011va} relies crucially on a maximal principal argument, which is invalidated by the presence of a gauge field.  

Fortunately, for boundary value problems with well-posed boundary conditions, the ellipticity of the equations guarantees that solutions are locally unique.  In particular, the solutions of the Einstein-Maxwell equations cannot be arbitrarily close to DeTurck solitons.  We should therefore be able to distinguish between DeTurck solitons and true solutions to Einstein-Maxwell by a careful monitoring of $\xi^a\xi_a\geq0$.

To solve the resulting PDEs, we employ Newton-Raphson iteration using pseudo-spectral collocation on a (possibly patched) Chebyshev grid.  Our patched grids are non-overlapping and formed using transfinite interpolation.  

This method is expected to have exponential convergence with increasing grid size if the metric functions are smooth.  While the equations for extremal horizons can sometimes yield highly non-analytic solutions, we expect our solutions to approach known, smooth extremal horizons.  We have checked that our solutions exhibit the expected exponential convergence of spectral methods, down to machine precision.  More specifically, we checked that the maximum value of the DeTurck norm $|\xi_{N}^2|_{\mathrm{max}}$ and the error in entropy $1-\mathcal S_{N-1}/\mathcal S_{N}$ decreases exponentially with increasing grid size $N$. 

\subsection{Background solutions with defects}\label{defects}

In this section, we detail the numerical construction of `background' solutions with defects on the boundary, but no black hole in the bulk.  As we mentioned earlier, the DeTurck method requires an appropriate choice of reference metric.  For this purpose, we just choose the $AdS_4$ metric as written in \eqref{eq:polar}, but with a new radial coordinate
\be\label{eq:neweta}
\eta = \frac{\bar\eta^2\sqrt{2-\bar\eta^4}}{1-\bar\eta^4}\;,
\ee
so that $\bar\eta\in(0,1)$.  Our metric ansatz is then
\begin{align}\label{eq:localisedansatz}
 \mathrm{d}s^2= \frac{L^2}{(1-\xi^2)^2}&\bigg[-\frac{\bar\eta^4(2-\bar\eta^4)}{(1-\bar\eta^4)^2}Q_1\mathrm{d}t^2+\frac{16\,Q_2\,\mathrm{d}\bar\eta^2}{\bar\eta^2(2-\bar\eta^4)^2(1-\bar\eta^4)^2}\\
 &\qquad\qquad+\frac{4\,Q_3}{2-\xi^2}\left(\mathrm{d}\xi-\frac{\xi\,Q_5\mathrm{d}\bar\eta}{\bar\eta}\right)^2+\xi^2(2-\xi^2)\,Q_4\,\mathrm{d}\phi^2\bigg]\,,
\end{align}
and for the gauge field, we choose
\be
A=L\, Q_6\;\dd t\;.
\ee
The $Q_i$ are functions of the coordinates $\bar\eta$ and $\xi$.  Written this way, the $AdS$ length scale $L$ drops out of our equations of motion.

Let us now discuss the boundary conditions.  At the conformal boundary $\xi=1$, we require the boundary metric to be conformal to Minkowski space, and for the gauge field to approach our profile $\mu(r)$.  The relationship between the coordinates $r$ and $\bar\eta$ on the boundary is given by the coordinate transformation \eqref{eq:transf} and \eqref{eq:neweta}.  That is, at the boundary we require
\be
Q_i(\bar\eta,\xi=1) = \left\{
     \begin{array}{ll}
       1 & : i =1,\ldots,4\\
       0 & : i=5\\
       \mu\left(\frac{1-\bar\eta^4}{\bar\eta^2\sqrt{2-\bar\eta^4}}\right) & : i=6\\
     \end{array}
   \right. \;,
\ee
Since $\bar\eta=1$ is the `origin' of the boundary metric, we require a similar condition there:
\be
Q_i(\bar\eta=1,\xi) = \left\{
     \begin{array}{ll}
       1 & : i =1,\ldots,4\\
       0 & : i=5\\
       \mu\left(0\right) & : i=6\\
     \end{array}
   \right. \;.
\ee
At the axis $\xi=0$, we require regularity.  This means
\begin{align}
\partial_\xi Q_i(\bar\eta,\xi=0) &= 0,\qquad\qquad\qquad :i\neq 4\nonumber \\
Q_4(\bar\eta,\xi=0)&=Q_3(\bar\eta,\xi=0)
\end{align}
There is an extremal horizon at $\bar\eta=0$ where we again require regularity:
\begin{align}
Q_1(\bar\eta=0,\xi)&=Q_2(\bar\eta=0,\xi)\nonumber\\
\partial_{\bar\eta} Q_i(\bar\eta=0,\xi) &= 0,\qquad\qquad\qquad :i=2,\ldots,4\nonumber \\
Q_i(\bar\eta=0,\xi)&=0,\qquad\qquad\qquad :i=5, 6
\end{align}
where we have assumed for consistency that $\lim_{r\rightarrow\infty}\mu(r)=0$.  Notice that our boundary condition at the extremal horizon does not fix the IR geometry there.  Even though the IR horizon in the reference metric is the Poincar\'e horizon, the solution is allowed to be something else, such as the point charge described in section \ref{sec:pointcharge}.  As mentioned in section \ref{sec:falloffs}, the IR geometry we find  depends upon the falloff of $\mu$.  

Since we are interested in seeing how various quantities change as we scale the chemical potential, we choose $\mu(r)=a\,p(r)$
for some fixed profile $p(r)$, and vary the number $a$.  We begin with a small $a$, where $AdS_4$ is a natural seed solution, then slowly increase $a$.  

After constructing these solutions, we compute the potential for charged (extremal) geodesics, and see if there is a value of $a$ at which this potential develops a minimum that is negative.  As we explained in section \ref{sec:geodesics}, this is the value of $a$ at which we expect static hovering black holes solutions to exist.  

\subsection{Hovering black holes}

In this section, we describe our construction of hovering extremal black holes in the `background' solutions computed in the previous section.  Compared to the `background' solutions, the construction of the solution with black holes has two major complications: an additional boundary in the integration domain, and the lack of an appropriate seed.  We will describe our approach to the first complication before discussing the second.

\begin{figure}[th]
\begin{center}
\includegraphics[width=.4\textwidth]{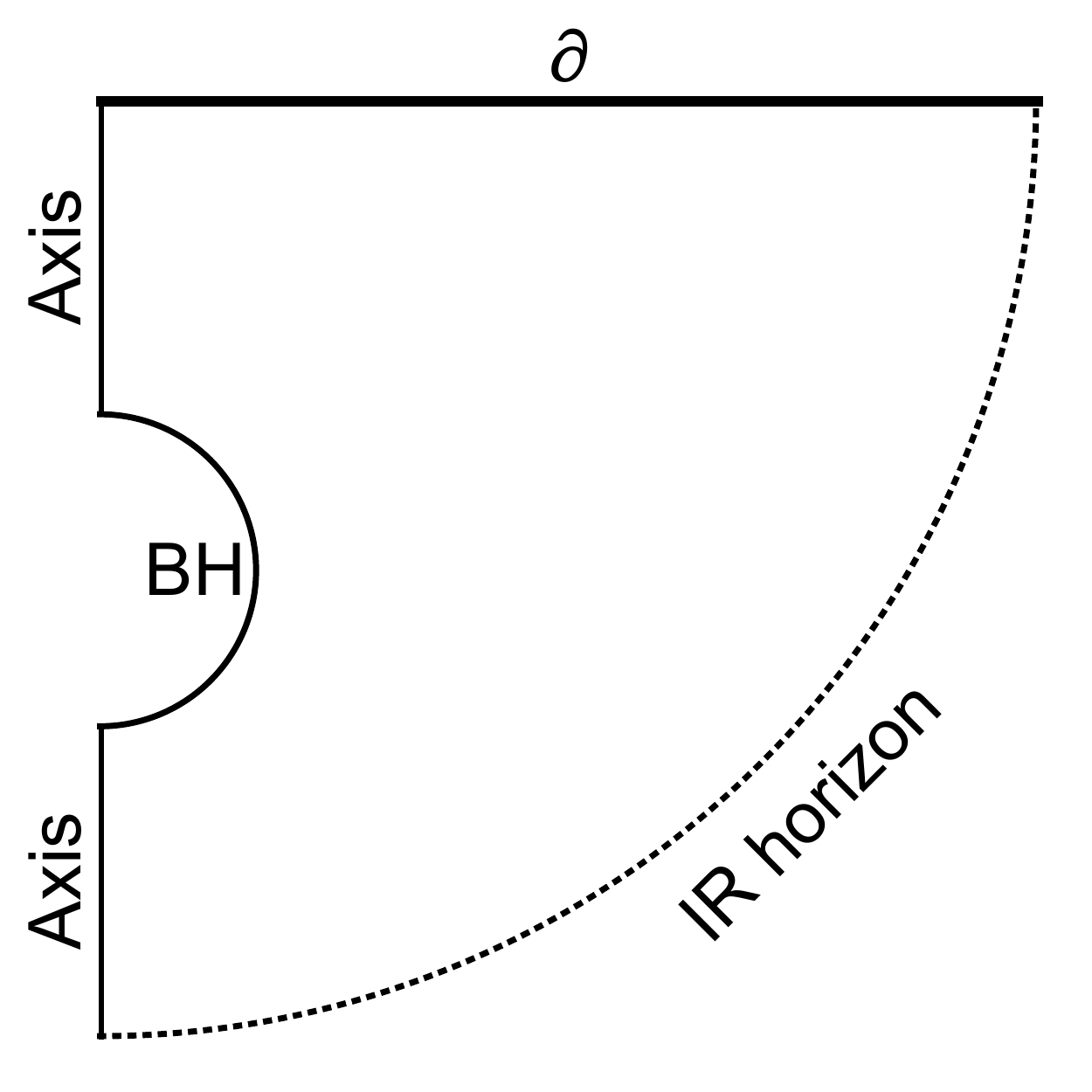}
\end{center}
\caption{Domain of integration for hovering black hole solutions.}\label{fig:domain}
\end{figure}

To begin, we must search for a reference metric that is compatible with our boundary conditions.  Our reference metric must have an extremal black hole horizon between two axes, an IR horizon, and the conformal boundary (see Fig.~\ref{fig:domain}).  Since there are five boundaries, two of which are horizons, we will choose to work in two different coordinate systems, each one adapted to one of the horizons.

\begin{figure}[th]
\begin{center}
\includegraphics[width=.4\textwidth]{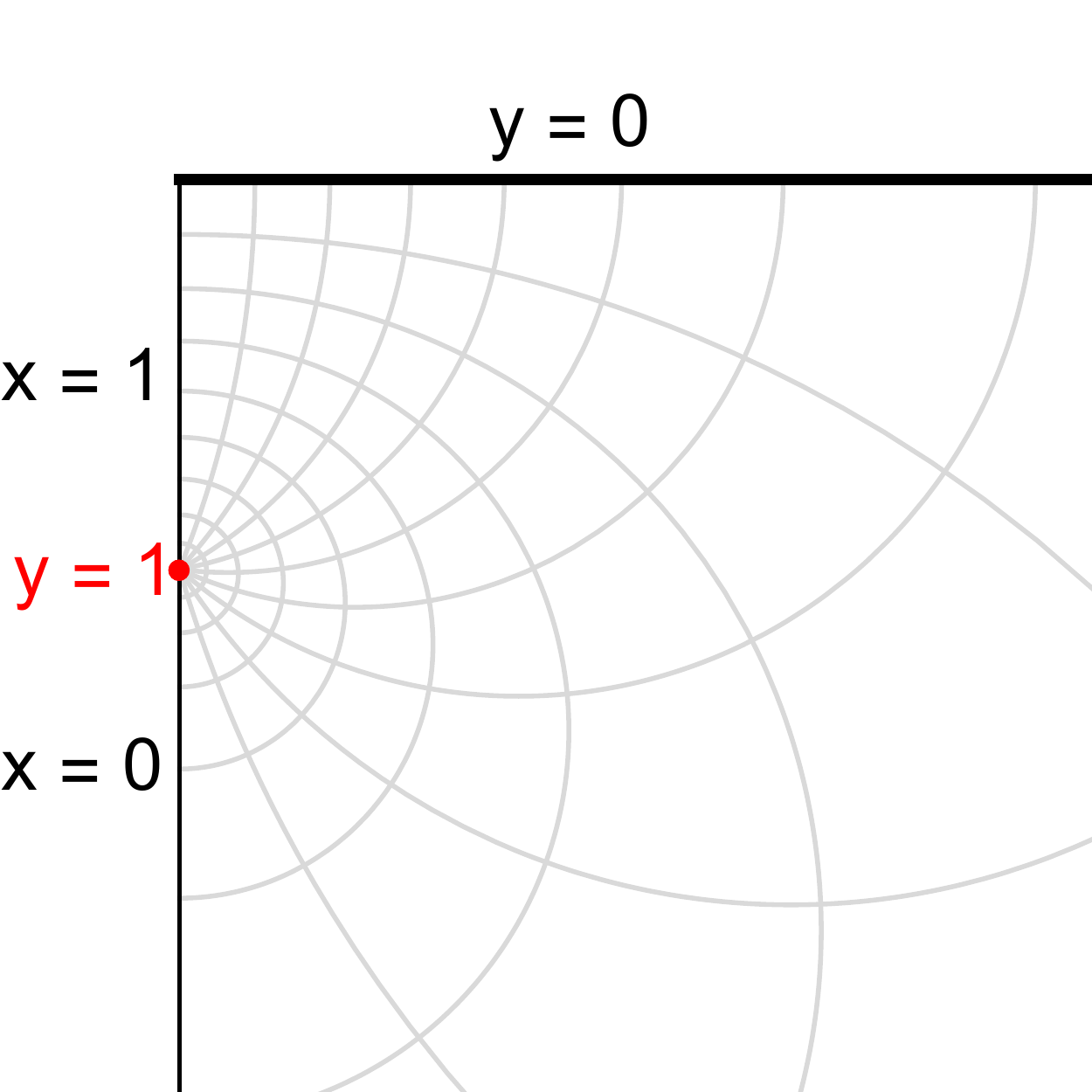}\hspace{1cm}
\includegraphics[width=.4\textwidth]{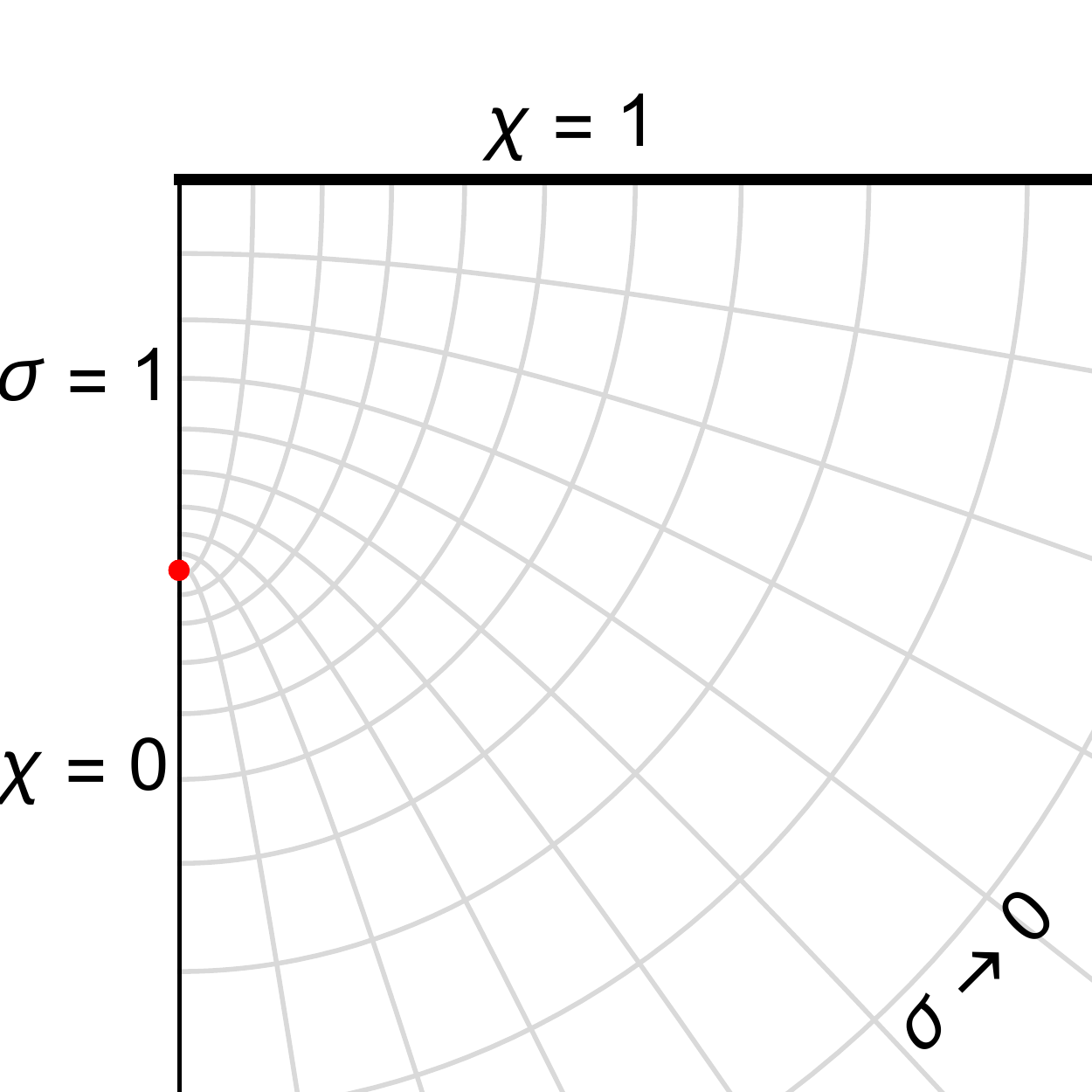}
\end{center}
\caption{Sketches for the bipolar (left) and elliptic (right) coordinate systems.}\label{fig:ellipticpolar}
\end{figure}

To aid in the construction of the reference metric, let us begin again with AdS in the usual Poincar\'e coordinates \eqref{eq:poincare}.  Now we perform a Cartesian to bipolar coordinate transformation
\be
r=\frac{x\sqrt{2-x^2}(1-x^2)(1-y^2)^2}{1-(1-x^2)^2(1-y^2)^2}\;,\qquad z=\frac{y\sqrt{2-y^2}}{1-(1-x^2)^2(1-y^2)^2}
\ee
to give us
\begin{align}\label{eq:bipolarads}
\dd s^2_{AdS}=\frac{L^2}{y^2(2-y^2)}&\bigg\{-g^2\dd t^2+\frac{4(1-y^2)^2\dd y^2}{2-y^2}\nonumber\\
&\qquad\qquad+(1-y^2)^4\left(\frac{4\dd x^2}{2-x^2}+x^2(2-x^2)(1-x^2)^2\dd \phi^2\right)\bigg\}\;,
\end{align}
where
\be
g=1-(1-x^2)^2(1-y^2)^2\;.
\ee
See the left panel of Fig.~\ref{fig:ellipticpolar} for a sketch of this coordinate system.  In this new coordinate system, the axis is split into $x=0$ and $x=1$ with a bipolar centre between them at $y=1$.  The boundary is now at $y=0$, and the entire Poincar\'e horizon has been mapped to the point $x=y=0$.

Based on this line element \eqref{eq:bipolarads}, we write down the following reference metric
\begin{align}\label{eq:refa}
\dd s^2_{\mathrm{ref}}=\frac{L^2}{y^2(2-y^2)}&\bigg\{-g^2(1-y^2)^4\dd t^2+\frac{4\dd y^2}{(1-y^2)^2(2-y^2)}\nonumber\\
&\qquad\qquad+\frac{4\dd x^2}{2-x^2}+x^2(2-x^2)(1-x^2)^2\dd \phi^2\bigg\}\;.
\end{align}
This metric \eqref{eq:refa} only differs from \eqref{eq:bipolarads} by a few factors of $(1-y^2)$ in several components.  As a result, there is now an extremal horizon at $y=1$ where there used to be the bipolar centre.  Furthermore, the geometry at $y=0$ is unchanged.  This means that we will still recover the Poincar\'e horizon at $x=y=0$.  We therefore have a line element that is well-suited to the hovering black hole horizon (at the `line' $y=1$), but ill-suited for the IR horizon (at the `point' $x=y=0$).  

To find a coordinate system better suited to the IR horizon, notice that in going from \eqref{eq:bipolarads} to \eqref{eq:refa}, we multiplied the $dx^2$ and $dy^2$ components by the same factor of $(1-y^2)^4$.  This means that these components are still conformal to bipolar coordinates (i.e. still conformally flat), and we can move to any orthogonal coordinate system of flat space.  Let us then move to elliptic coordinates:
\be\label{eq:atob}
x=\sqrt{1-\frac{1-\sigma^4}{\sqrt{1-(1-\chi^2)^2\sigma^4(2-\sigma^4)}}}\;,\qquad y=\sqrt{1-\sqrt{1-(1-\chi^2)^2\sigma^4(2-\sigma^4)}}\;,
\ee
which gives
\be\label{eq:refb}
\dd s^2_{\mathrm{ref}}=\frac{L^2}{(1-\chi^2)^2}\bigg\{\sigma^4(2-\sigma^4)h^2\dd t^2+\frac{16\dd\sigma^2}{\sigma^2(2-\sigma^4)^2h}+\frac{4\dd\chi^2}{(2-\chi^2)h}+\frac{\chi^2(2-\chi^2)(1-\sigma^4)^2\dd\phi^2}{h^2}\bigg\}\;,
\ee
where
\be
h=1-(1-\chi^2)^2\sigma^4(2-\sigma^4)\;.
\ee
See the right panel of Fig.~\ref{fig:ellipticpolar} for a sketch of this coordinate system.  In these `elliptic' coordinates, the IR horizon is at $\sigma=0$, the boundary is at $\chi=1$, the two axes are at $\chi=0$ and $\sigma=1$, and the black hole horizon is at the point $\sigma=1$, $\chi=0$.  This coordinate system is better adapted to the IR horizon, but not the black hole horizon.  

To summarize, we have an appropriate reference metric described by the two coordinate systems \eqref{eq:refa} and \eqref{eq:refb}, and each boundary of our integration domain is well described in at least one of these coordinates.  The map between the coordinates is given by \eqref{eq:atob}.

Now let us write down a metric ansatz.  In bipolar $(x,y)$ coordinates, we have
\begin{align}\label{eq:ansatza}
\dd s^2=\frac{L^2}{y^2(2-y^2)}&\bigg\{-g^2(1-y^2)^4F_1\dd t^2+\frac{4F_2\dd y^2}{(1-y^2)^2(2-y^2)}\nonumber\\
&\qquad\qquad+\frac{4F_3}{2-x^2}\left(\dd x- \frac{x(2-x^2)F_5\dd y}{(1-y^2)g}\right)^2+x^2(2-x^2)(1-x^2)^2F_4\dd \phi^2\bigg\}\;,
\end{align}
while in elliptic $(\sigma,\chi)$ coordinates, we have
\begin{align}\label{eq:ansatzb}
\dd s^2=\frac{L^2}{(1-\chi^2)^2}&\bigg\{\sigma^4(2-\sigma^4)h^2G_1\dd t^2+\frac{16G_2\dd\sigma^2}{\sigma^2(2-\sigma^4)^2h}\nonumber\\
&\qquad+\frac{4G_3}{(2-\chi^2)h}\left(\dd\chi-\frac{2\chi(2-\chi^2)G_5\dd\sigma}{\sigma(2-\sigma^4)h}\right)^2+\frac{\chi^2(2-\chi^2)(1-\sigma^4)^2G_4\dd\phi^2}{h^2}\bigg\}\;.
\end{align}
As for the gauge field, we choose
\be
A=L\,F_6\,\dd t=L\,G_6\,\dd t\;.
\ee
We treat $F_i$ as functions of $(x,y)$ and $G_i$ as functions of $(\sigma,\chi)$.  

The boundary conditions are similar to those for the `background' defect solution in section \ref{defects}.  At the boundary $y=0$ or $\chi=1$, we have
\be
F_i(x,y=0) = \left\{
     \begin{array}{ll}
       1 & : i =1,\ldots,4\\
       0 & : i=5\\
       \mu\left(\frac{1-x^2}{x\sqrt{2-x^2}}\right) & : i=6\\
     \end{array}
   \right. \;,\quad
G_i(\sigma,\chi=1) = \left\{
     \begin{array}{ll}
       1 & : i =1,\ldots,4\\
       0 & : i=5\\
       \mu\left(\frac{1-\sigma^4}{\sigma^2\sqrt{2-\sigma^4}}\right) & : i=6\\
     \end{array}
   \right.\;,
\ee
The remaining boundary conditions are imposed by regularity.  At one of the axes, we have
\begin{align}
\partial_x F_i(x=0,y) &= 0,\qquad\qquad\qquad :i\neq 4\nonumber \\
F_4(x=0,y)&=F_3(x=0,y)\nonumber\\
\partial_\xi G_i(\sigma,\chi=0) &= 0,\qquad\qquad\qquad :i\neq 4\nonumber \\
G_4(\sigma,\chi=0)&=G_3(\sigma,\chi=0)\;.
\end{align}
At the other axis,
\begin{align}
\partial_x F_i(x=1,y) &= 0,\qquad\qquad\qquad :i\neq 4, 5\nonumber \\
F_4(x=1,y)&=F_3(x=0,y)\nonumber\\
F_5(x=1,y)&=0\nonumber\\
\partial_\chi G_i(\sigma=1,\chi) &= 0,\qquad\qquad\qquad :i\neq 4,5\nonumber \\
G_4(\sigma=1,\chi)&=G_3(\eta=1,\chi)\nonumber\\
G_5(\sigma=1,\chi)&=0\;.
\end{align}
We restrict ourselves to imposing boundary conditions on the black hole horizon in bipolar coordinates, and the IR horizon in elliptic coordinates.  These conditions are
\begin{align}
\partial_y F_i(x,y=1) &= 0,\qquad\qquad\qquad :i= 2,\ldots,4,\nonumber \\
F_1(x,y=1)&=F_2(x,y=1)\nonumber\\
F_5(x,y=1)&=0\nonumber\\
F_6(x,y=1)&=0\;,
\end{align}
for the black hole horizon, and
\begin{align}
\partial_y G_i(\sigma=0,\chi) &= 0,\qquad\qquad\qquad :i= 2,\ldots,4,\nonumber \\
G_1(\sigma=0,\chi)&=G_2(\sigma=0,\chi)\nonumber\\
G_5(\sigma=0,\chi)&=0\nonumber\\
G_6(\sigma=0,\chi)&=0\;,
\end{align}
for the IR horizon. 

In practice, we partition the integration domain into two non-overlapping `patches', one in each coordinate system.  We place a grid on each patch using transfinite interpolation, then solve the system subject to the above boundary conditions.  Since we have two non-overlapping patches, in addition to the conditions above, we have additional patching conditions on any (artificial) patch boundaries where we impose that the two line elements \eqref{eq:ansatza} and \eqref{eq:ansatzb} and their first derivatives are equivalent under the transformation \eqref{eq:atob}.  We choose our patch boundary to extend from the point $(x=1,y=0)$ or $(\sigma=1,\chi=1)$ to somewhere along the line $x=0$ or $\chi=0$.  

In order to find a solution using our methods (Newton-Raphson), we require a good seed solution.  Unfortunately, the solutions we are looking for are not close to any previously known solution to the Einstein-Maxwell equations, so there are no readily available seeds.  To remedy this, consider the following equations
\be\label{eq:deltaEOM}
G^H_{\mu\nu}[g,A]-\delta\,G^H_{\mu\nu}[\bar g,A]=0\;,\qquad \nabla_\mu F^{\mu\nu}=0\;,
\ee 
where $\delta$ is a constant, $g$ is the metric, $\bar g$ is the reference metric, and we are now working in a coordinate basis.  By construction, $\delta=1$, $g=\bar g$, and $A=0$ is a solution to the above equations of motion, and $\delta=0$ recovers the same equations of motion as \eqref{eq:EMD}.  Furthermore, \eqref{eq:deltaEOM} has the same differential operator as \eqref{eq:EMD}, so they are also Elliptic equations.  The new equations are also consistent with the boundary conditions we have outlined above.  

We can therefore attempt to find solutions to \eqref{eq:EMD} using the following procedure.  Begin with $\delta=1$, $g=\bar g$, and $A=0$, which is a solution to \eqref{eq:deltaEOM}.  Now continue to solve \eqref{eq:deltaEOM} by slowly increasing the amplitude $a$ until it exceeds the value at which black holes are expected to form.  (This value is determined by an analysis of the `background' solutions whose construction we described in section \ref{defects}.)  Then we slowly decrease $\delta$ until we reach $\delta=0$, where we would have a solution to \eqref{eq:EMD}.

\section{Numerical Results}
\begin{figure}[th]
\begin{center}
\includegraphics[width=.7\textwidth]{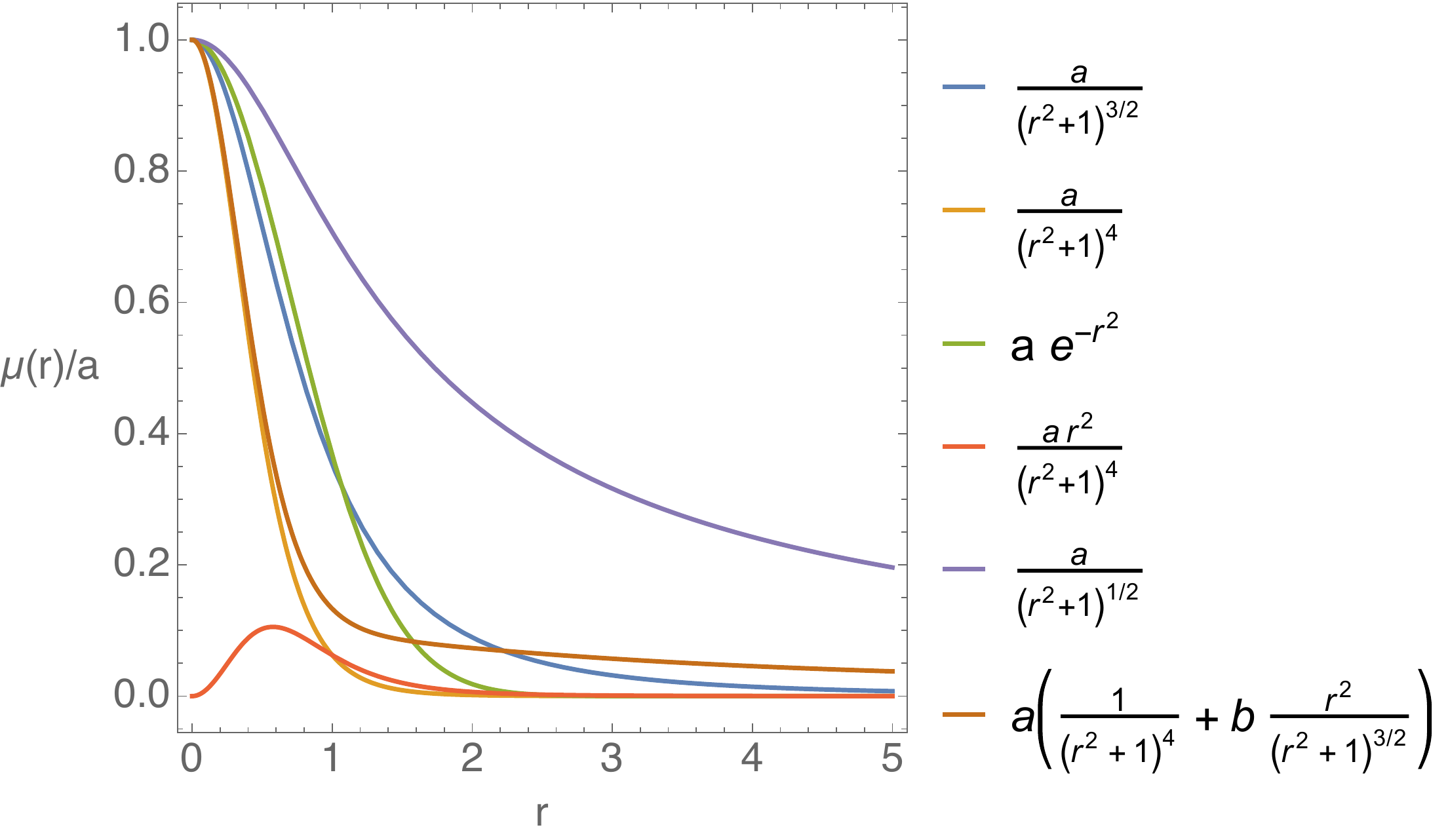}
\end{center}
\caption{\label{fig:pro}Sketches for the profiles in Eq.~(\ref{eq:profiles}).}
\end{figure}

Let us now discuss the results of our numerical construction which was outlined above.  We have studied a number of profiles for the chemical potential:

\begin{subequations}
\begin{equation}
\begin{array}{rl}
 \mu_{I_1}(r)&\displaystyle=\frac{a}{\left(\frac{r^2}{\ell^2}+1\right)^{3/2}}
 \\
\mu_{I_2}(r)&\displaystyle=\frac{a}{\left(\frac{r^2}{\ell^2}+1\right)^4}
\\
\mu_{I_3}(r)&\displaystyle=a\,e^{-\frac{r^2}{\ell^2}}
\\
\mu_{I_4}(r)&\displaystyle=\frac{a\,r^2}{\ell^2\,\left(\frac{r^2}{\ell^2}+1\right)^4}
\end{array}
\label{eq:profilesi}
\end{equation}
\begin{equation}
\begin{array}{rl}
\mu_{M_1}(r)&\displaystyle=\frac{a}{\left(\frac{r^2}{\ell^2}+1\right)^{1/2}}
\\
\vspace{-0.5cm}
\\
\mu_{M_2}(r)&\displaystyle=a\left[\frac{1}{\left(\frac{r^2}{\ell^2}+1\right)^4}+\frac{b \,r^2}{\ell^2(\frac{r^2}{\ell^2}+1)^{3/2}}\right]\;.
\end{array}
\label{eq:profilesm}
\end{equation}
\label{eq:profiles}
\end{subequations}
In the profile $M_2$, there is an additional parameter $b$ which we keep fixed when varying $a$.  Note that these profiles include a length scale $\ell$.  Due to the conformal symmetry of the UV theory, our results only depend upon the product $a \ell$.  We therefore henceforth set $\ell=1$.  These profiles are plotted in Fig.~\ref{fig:pro}.

We can divide the profiles above into two groups depending on their falloffs.  The first four (profiles $I_1$ to $I_4$) have a falloff faster than $1/r$, and so are `irrelevant' profiles.  The final two (profiles $M_1$ and $M_2$) fall as $1/r$ and so are `marginal' profiles.  Absent from the present section are any profiles with a falloff slower than $1/r$ (i.e. `relevant' profiles).  Our numerics do not seem to allow us to study such profiles at zero temperature.    

\subsection{Background solutions}

 We first consider the gravitational solutions without a black hole.  We begin by comparing our numerical results to perturbation theory and our expectations in \ref{sec:falloffs}, then discuss the existence of a maximum amplitude for these solutions, and analyze the effective potential for the static orbits of extremal charged particles.  

\begin{figure}[th]
\begin{center}
\includegraphics[width=\textwidth]{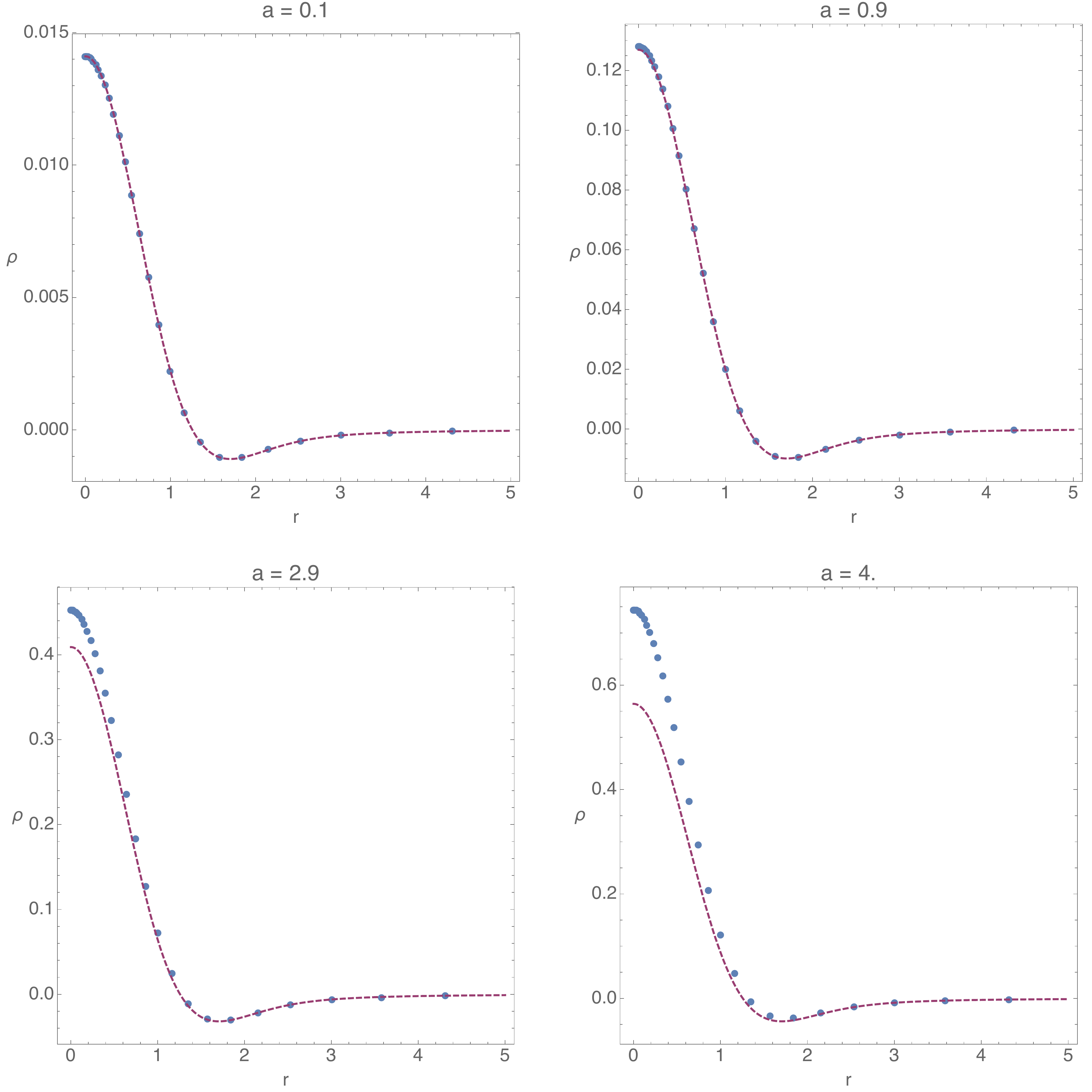}
\end{center}
\caption{\label{fig:charge}Charge density as a function of the boundary radial coordinate $r$ for profile $I_3$ defined in Eq.~(\ref{eq:profilesi}): the purple dashed line indicates the perturbative prediction whereas the blue dots represent our numerical data.}
\end{figure}

In the irrelevant case, profiles with small amplitudes can be studied perturbatively about $AdS_4$.  The details for this perturbative calculation are in Appendix \ref{app:1}.  The charge density $\rho(r)$ can be extracted from the behaviour of the gauge field near the boundary.  According to our perturbative analysis, the charge density, $\rho(r)$ for our irrelevant profiles has a $1/r^3$ falloff\footnote{For any irrelevant profile that decays slower than $1/r^2$, the charge density instead decays as $\mu(r)/r$ at large $r$.  A profile with precisely $1/r^2$ falloff has a charge density decaying as $\log r/r^3$.  For simplicity, we did not consider irrelevant profiles with slower than $1/r^2$ falloff.}.  Our irrelevant profiles agree with this perturbative analysis.  Fig.~\ref{fig:charge} contains four panels with plots of the charge density for several values of the amplitude $a$ for one of our profiles. The purple dashed line is the analytic prediction from the perturbation theory.  The agreement at large $r$ is good, confirming the expected falloff of the charge density.

\begin{figure}[th]
\begin{center}
\includegraphics[width=0.5\textwidth]{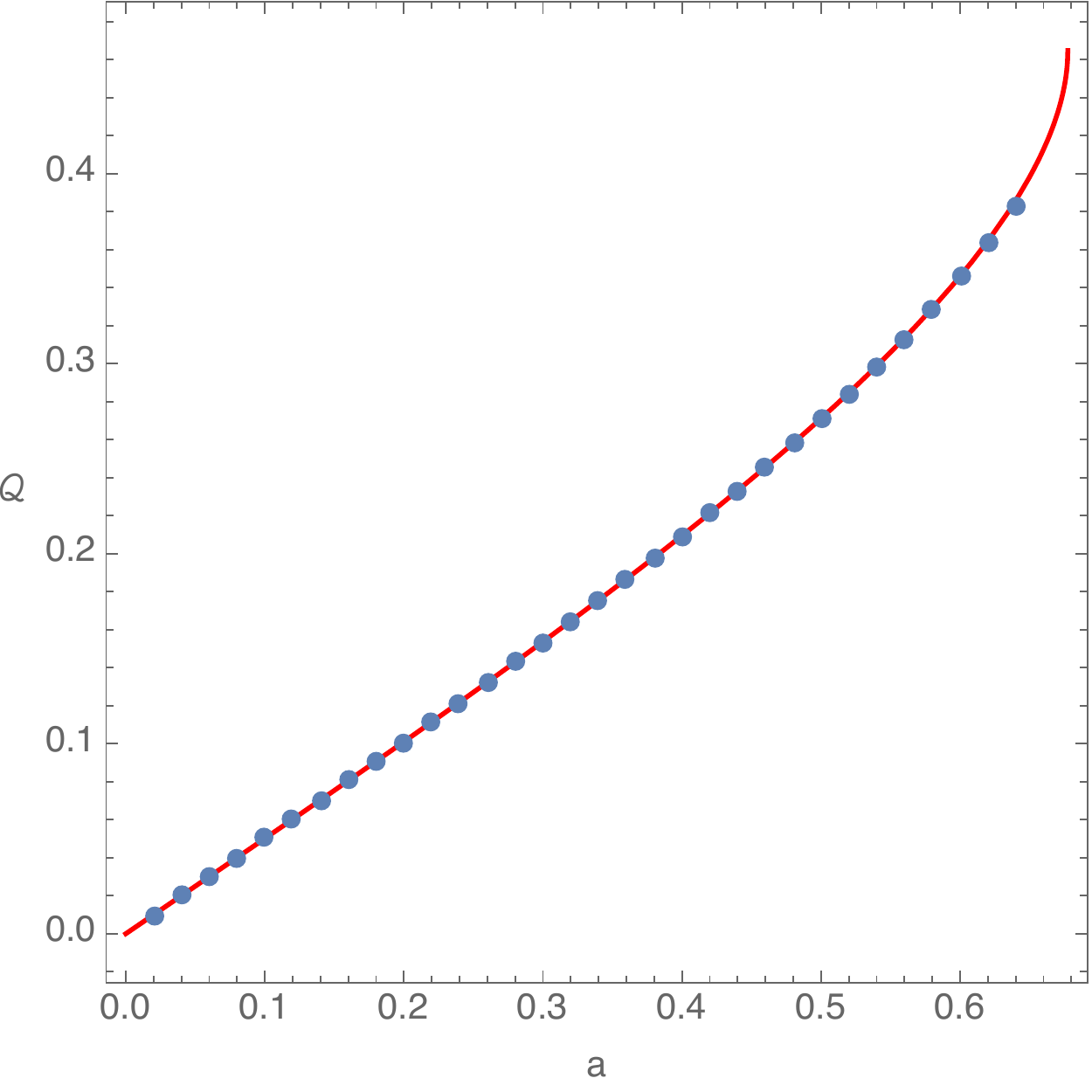}
\end{center}
\caption{\label{fig:marginalrho}Total charge as a function of amplitude. The red line is the value for the point charge (up to $a_{\max}$), and the dots are numerical data for the profile $M_1$.}
\end{figure}

As we mentioned in section \ref{sec:pointcharge}, the charge density of the point charge solution is located at the origin in a delta function.  In our marginal profiles, there is no longer such a delta function since the chemical potential is smooth.  As an additional numerical check, we verify that the total charge of our marginal profiles matches that of the point charge with the same $1/r$ falloff, as can be seen in Fig.~\ref{fig:marginalrho}.

\begin{figure}[th]
\begin{center}
\includegraphics[width=0.9\textwidth]{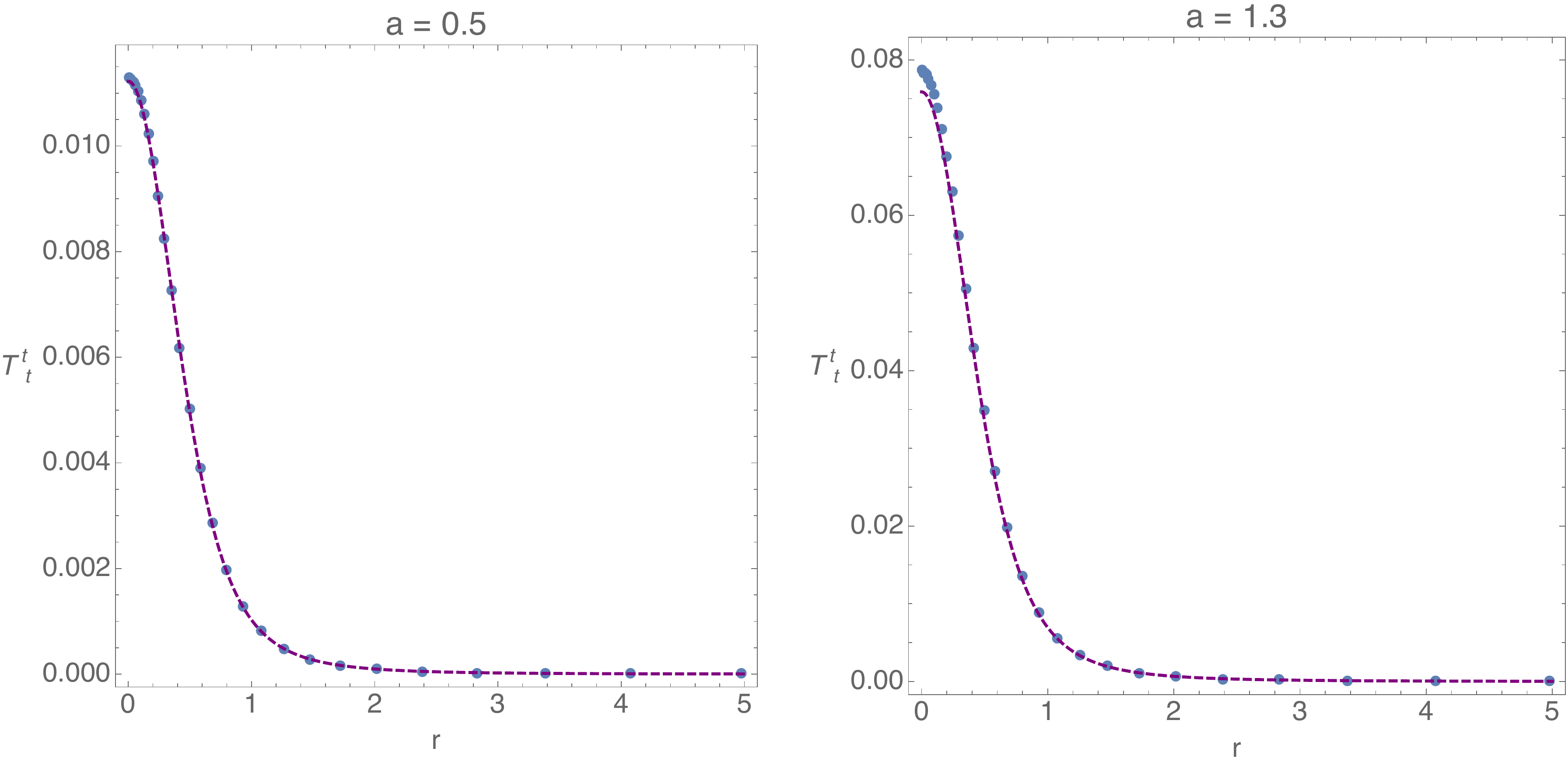}
\end{center}
\caption{\label{fig:stress}Stress energy tensor as a function of the boundary radial coordinate $r$ for $I_1$ defined in Eq.~(\ref{eq:profilesi}): the purple dashed line indicates the perturbative prediction whereas the blue dots represent our numerical data.}
\end{figure}

\begin{figure}[th]
\begin{center}
\includegraphics[width=0.5\textwidth]{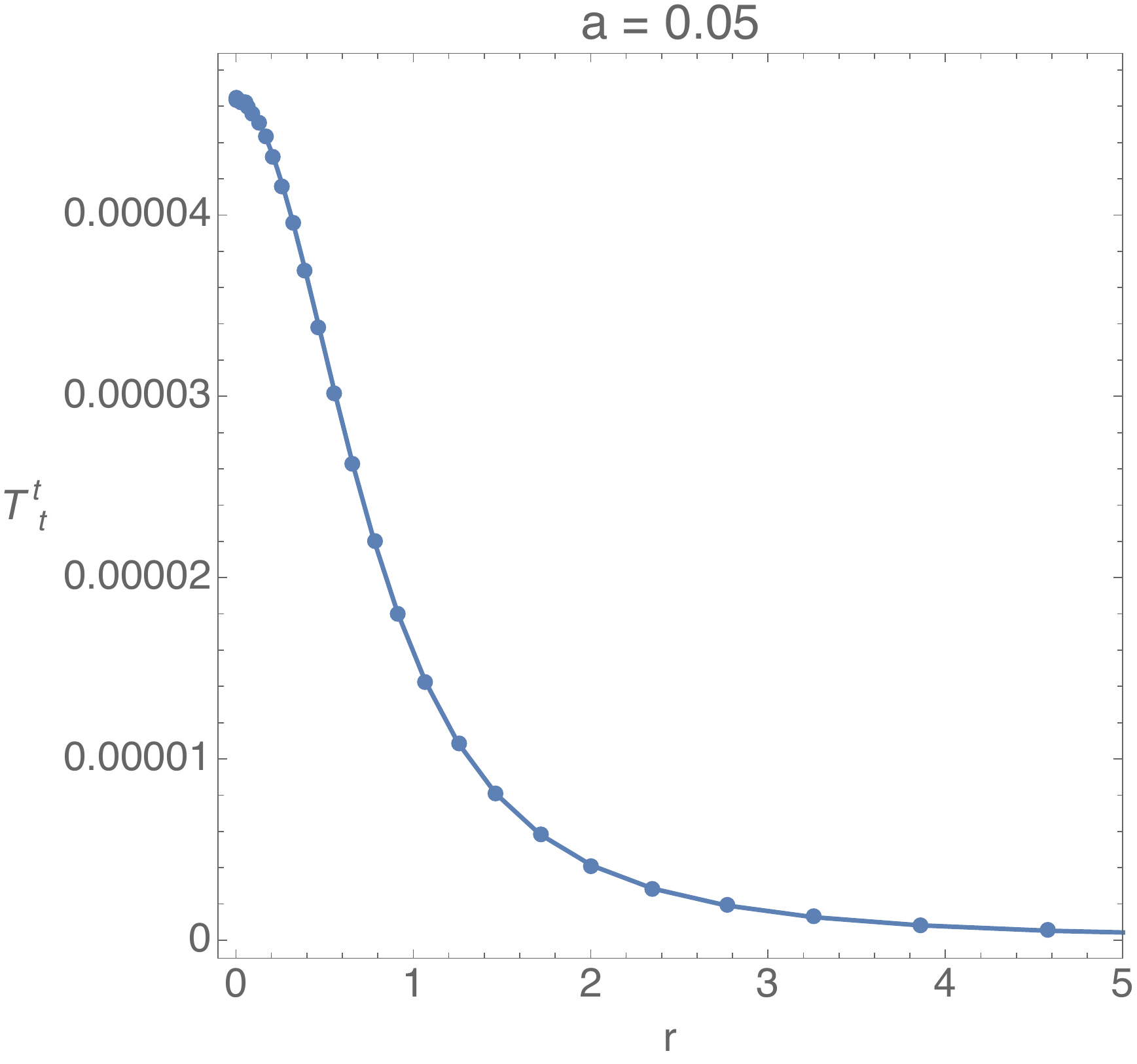}
\end{center}
\caption{\label{fig:stressm}Stress energy tensor as a function of the boundary radial coordinate $r$ for $M_1$ defined in Eq.~(\ref{eq:profilesm}).}
\end{figure}

We can repeat this calculation for the energy density $T^t_t$, which can be extracted using a standard holographic renormalisation procedure \cite{Balasubramanian:1999re,deHaro:2000xn}. In this case we expect the large $r$ behaviour of our profiles to have a $1/r^5$ falloff \eqref{energyfalloff} possibly modified by $\ln r$ terms\footnote{This again depends on choosing irrelevant profiles that decay faster than $r^{-2}$.}. We again find good quantitative agreement between numerics and perturbation theory. Two typical runs for the first profile in Eq.~(\ref{eq:profilesi}) are shown in Fig.~\ref{fig:stress}. For this profile, the decay extracted at large $r$ is compatible with $r^{-5}\log r$.  For completeness, we also include the stress tensor for one of our marginal profiles in Fig. ~\ref{fig:stressm}.  The falloff of this stress tensor goes as $1/r^3$, consistent with our discussion in section \ref{sec:falloffs}.

\begin{figure}[th]
\begin{center}
\includegraphics[width=0.5\textwidth]{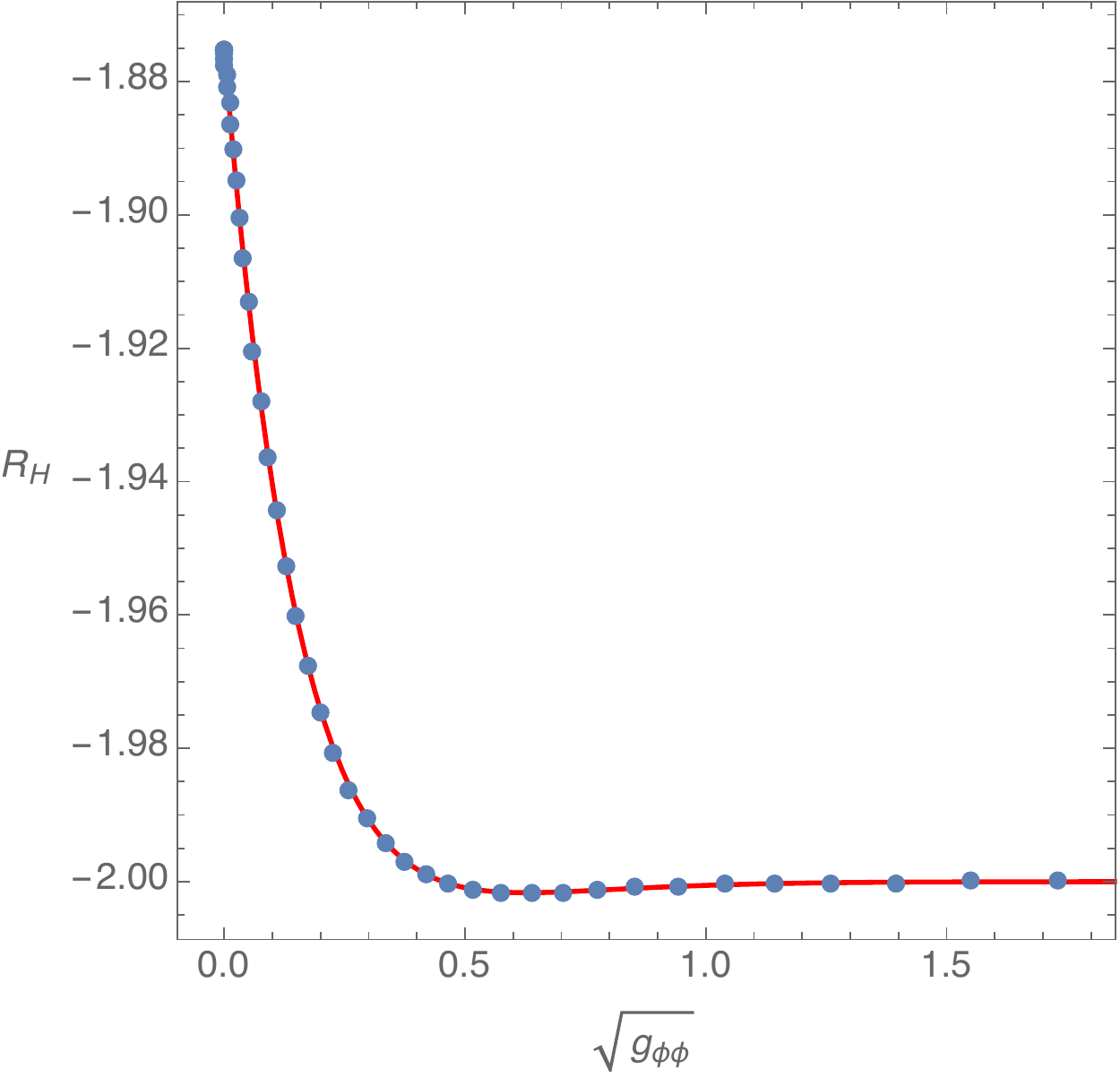}
\end{center}
\caption{\label{fig:solenoidir}Ricci scalar the of induced IR horizon geometry a function of $\sqrt{g_{\phi\phi}}$ on the horizon.  The red line is the analytic result for the point charge, the dots are numerical data.  Here, we use profile $M_1$ with $a=0.2$.}
\end{figure}

We also note that we did not input the IR geometry into our code.  So as an additional check of our numerics, we can verify that the IR geometry is the same as the Poincar\'e horizon for the irrelevant profiles and the same as the point charge \eqref{pcharg} for the marginal profiles.   This can be seen, for instance, by computing the Ricci scalar of the induced horizon geometry as a function of $\sqrt{g_{\phi\phi}}$ on the horizon.  This is plotted in Fig.~\ref{fig:solenoidir} for one of the marginal profiles.

\begin{figure}[th]
\begin{center}
\includegraphics[width=.5\textwidth]{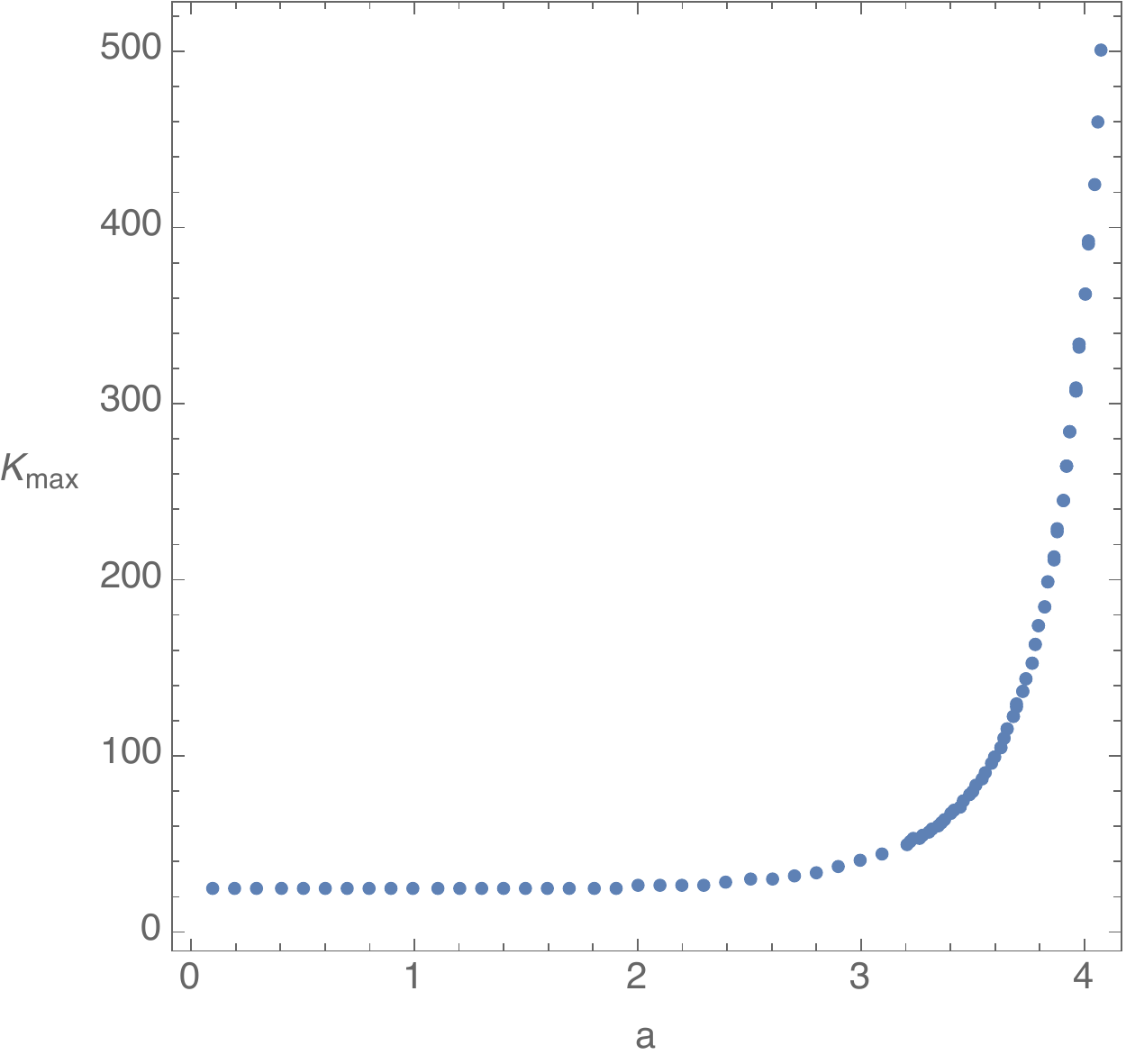}
\end{center}
\caption{\label{fig:kre}Maximum value of the Kretschmann scalar for one of the irrelevant profiles as a function of the amplitude. For $a = 0$, we recover the AdS$_4$ result $24/L^4$.}
\end{figure}

For each of the profiles, we are unable to find background solutions with an amplitude above some (profile dependent) value $a_{\max}$.  In the irrelevant case, these solutions appear to be becoming singular.  Fig.~\ref{fig:kre} shows the maximum value of the Kretschmann scalar as a function of the amplitude for one of the irrelevant profiles.  The Kretschmann scalar begins at the $AdS_4$ value of $24/L^4$ for small amplitudes and increases rapidly near $a\approx 4$.   In the point charge solution in section \ref{sec:pointcharge}, there is also a maximum value $a_{\max}\approx 0.6675$.  We were unable to extend our marginal profiles above this value, nor were we able to construct a second branch of solutions.  This is consistent with our discussion in section \ref{sec:pointcharge}.  Here, the point charge solutions and the solutions with a marginal profile do not appear to be singular.  

Now we compute the effective potential $\mathcal V$ for timelike static orbits of (extremal) charged test particles.  Rather than use a specific coordinate system, we opt to plot the potential as a function of the thermal length $\ell_T \equiv \sqrt{-g_{tt}}$.  Note that the Poincar\'e horizon has thermal length $\ell_T =0$.  

\begin{figure}[th]
\begin{center}
\includegraphics[width=.6\textwidth]{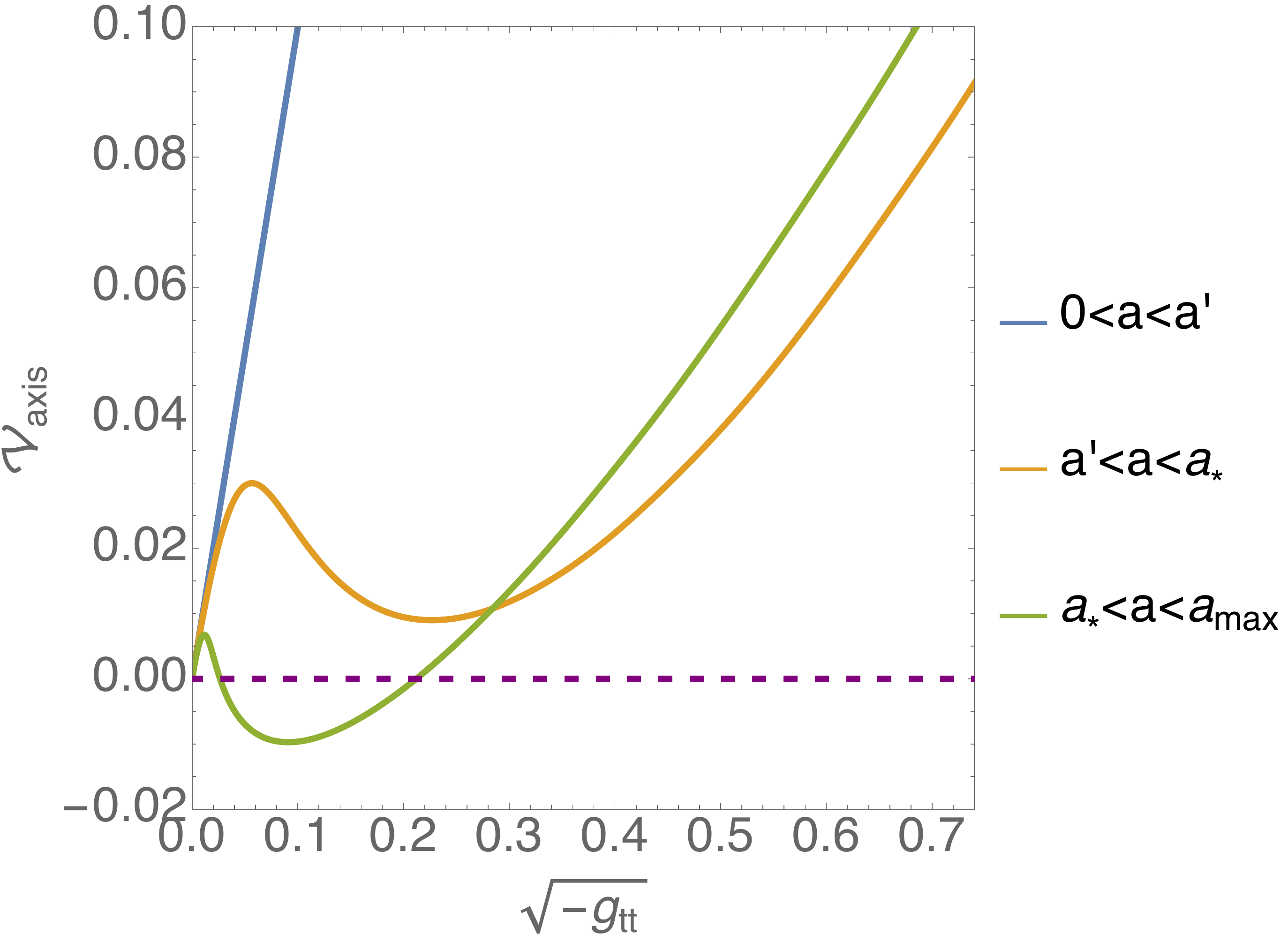}
\end{center}
\caption{\label{fig:orbits}Plots of the potential Eq.~(\ref{eq:potgeo}) along the axis for a representative profile.}
\end{figure}

For profiles with amplitudes in the range $0<a<a_{max}$, we typically find three different regimes for the qualitative behaviour of $\mathcal V$.  For $0<a<a^\prime$, $\mathcal{V}$ does not have any extrema. For $a^\prime<a<a_{\star}$, there is a local maximum and a local minimum in $\mathcal{V}$, but the minimum is above zero. For $a_{\star}<a<a_{\max}$, there is still a maximum and minimum, but the minimum is now below zero.   All of these regimes can be seen in Fig.~\ref{fig:orbits} where we plot the potential along the axis for a representative profile.

\begin{figure}[th]
\begin{center}
\includegraphics[width=.7\textwidth]{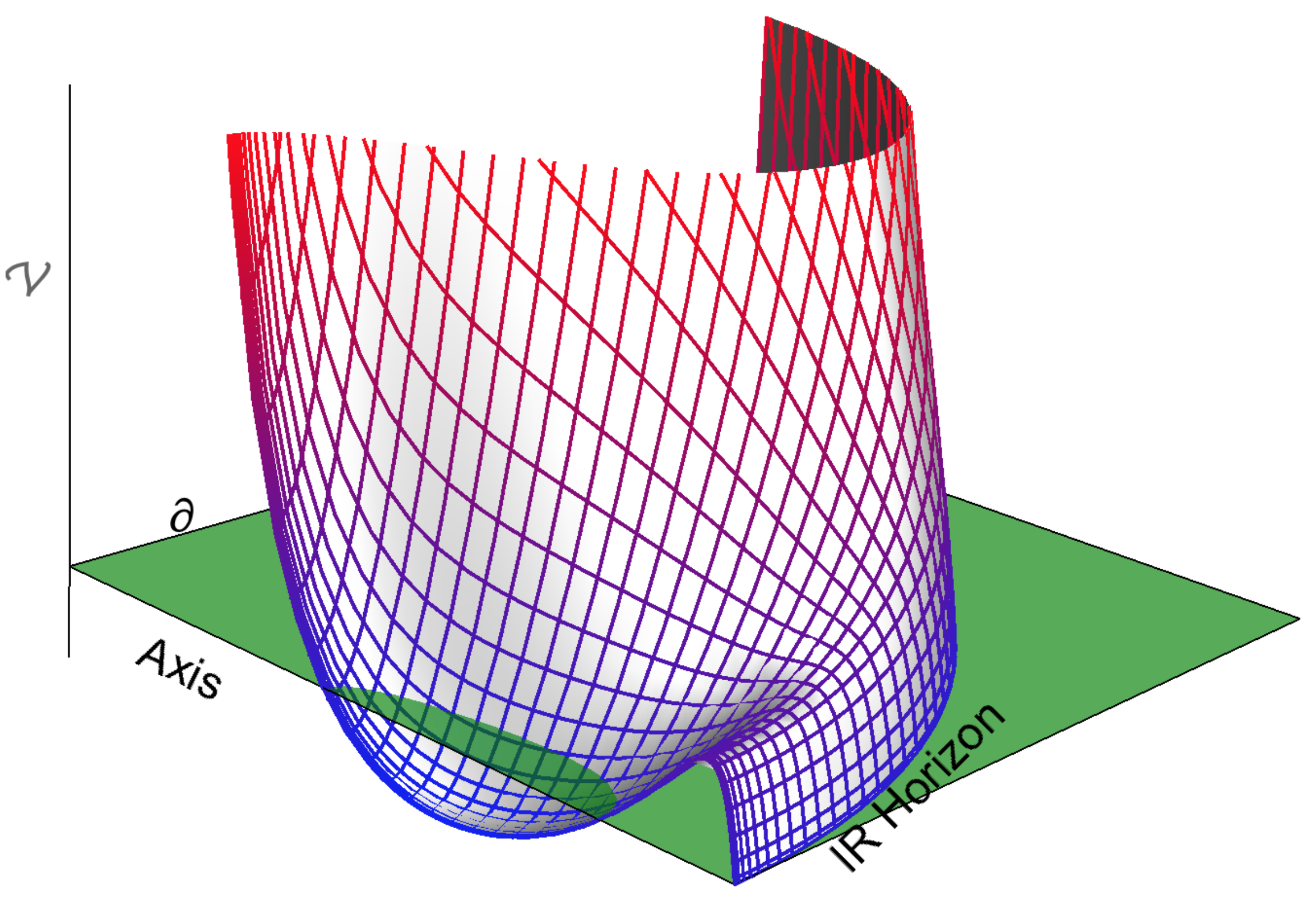}
\end{center}
\caption{\label{fig:potential3d}Shape of the potential Eq.~(\ref{eq:potgeo}) for a representative profile with $a_{\star}<a<a_{\mathrm{max}}$.  The green plane represents $\mathcal V=0$.  The potential is zero on the IR horizon (here mapped to a quarter circle), and diverges as it approaches the boundary.  There is a  global minimum on the axis.}
\end{figure}

The potential (including off the axis) can be visualised in Fig.~\ref{fig:potential3d}.  There, we have mapped our coordinate system into a quarter disk with the quarter circle representing the IR horizon and the two sides representing the symmetry axis and the boundary.  This is not gauge invariant, but nevertheless provides a visual aid for the shape of the potential.  One can clearly see that the minimum along the axis is indeed a true minimum and not a saddle point.  

The amplitudes $a'$ and $a_{\max}$ thus represent critical amplitudes.  As we have explained in section \ref{sec:geodesics}, $a'$ is the critical amplitude above which a small extremal charged particle is stable in the geodesic approximation. Above $a_{\star}$, we expect hovering black hole solutions to exist. 

\begin{figure}[th]
\begin{center}
\includegraphics[width=.6\textwidth]{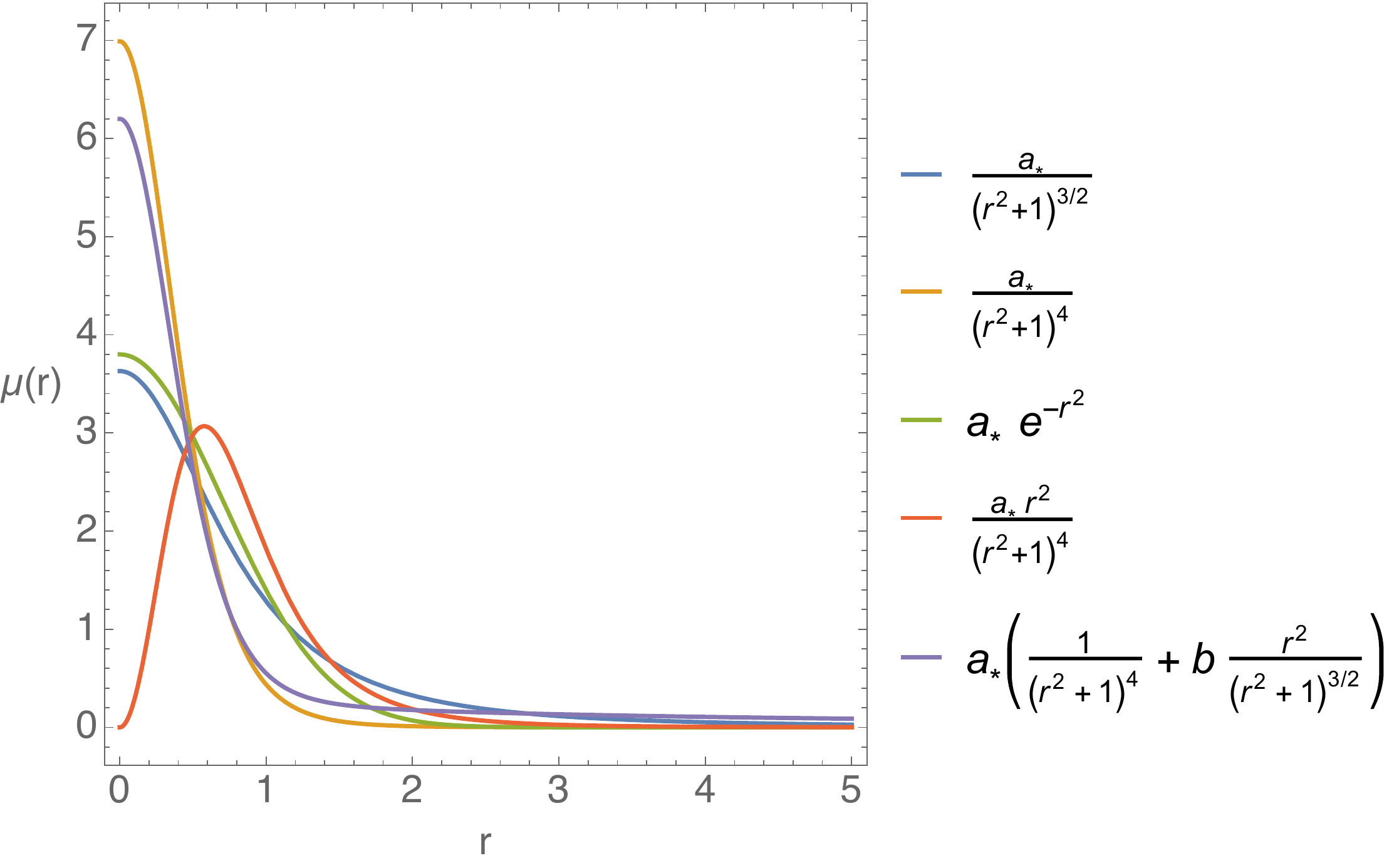}
\end{center}
\caption{\label{fig:pro1}Sketches for the profiles Eq.~(\ref{eq:profiles}) for $a = a_{\star}$.  In this plot, we set $b=0.075$.}
\end{figure}

In table \ref{tab:1} we show the values for $a_{\star}$ for all of the profiles in Eq.~(\ref{eq:profiles}). In addition, we also plot in Fig.~\ref{fig:pro1} what the several profiles look like for $a = a_{\star}$.  As one can see, these curves differ significantly from one another.  We have attempted to search for possible physical quantities at $a=a_{\star}$ that might give a universal quantity or property for the black hole nucleation. These include: the total area under the profiles, the total enthalpy $\int_0^{+\infty} \mu(r)\rho(r)r dr$, the $r^{-3}$ coefficient of the large $r$ expansion of $\rho(r)$, the total energy and the Gibbs free energy. None of these show any universality.

Interestingly, unlike the irrelevant profiles we have tried, the marginal profiles do not necessarily have such a critical $a'$ or $a_{\star}$ below $a_{\max}$.  That is, no extrema develop.  This is not true for all marginal profiles since profile $M_2$ yields these critical amplitudes $a'$ and $a_{\star}$ for sufficiently small values of $b$.
  
  \begin{table}
  \begin{tabular}{|c|c|c|c|c|c|c|c|}
  \hline  \hline
  & $I_1$ & $I_2$ & $I_3$ & $I_4$ &$M_1$ &$M_2 (b=0.075)$
  \\
  \hline
 $a_{\star}$ & 3.63 & 6.99 & 3.8 & 29.09 & N/A & 6.20\\ 
    \hline \hline
  \end{tabular}
  \caption{ \label{tab:1}The critical value at which hovering black holes form for our various profiles in Eq.~(\ref{eq:profiles}).}
\end{table}

\subsection{Hovering black holes}

Now let us discuss our numerical results for the hovering solutions.  As explained earlier in this section and in section \ref{sec:geodesics}, all of these solutions have an amplitude $a>a_{\star}$, where $a_{\star}$ is the critical amplitude above which the effective potential for extremal charged geodesics develops a minimum below zero.  Since profile $M_1$ does not contain this critical value, it is omitted from this discussion.  Here, we fix the parameter $b=0.075$ in the profile $M_2$ which does have such an $a_{\star}$.  

Note that we have found hovering black holes for both marginal and irrelevant profiles.  These profiles yield different IR geometries.  As in the background case, we again verified here that the IR geometry approaches the Poincar\'e horizon for irrelevant profiles and that of the point charge for marginal profiles.  We also verified that the total charge is zero in the irrelevant case.  In the marginal case, we confirmed that the integral of the boundary charge density is equal to the sum of the charge on the black hole and on the IR horizon.

\begin{figure}[th]
\begin{center}
\includegraphics[width=.4\textwidth]{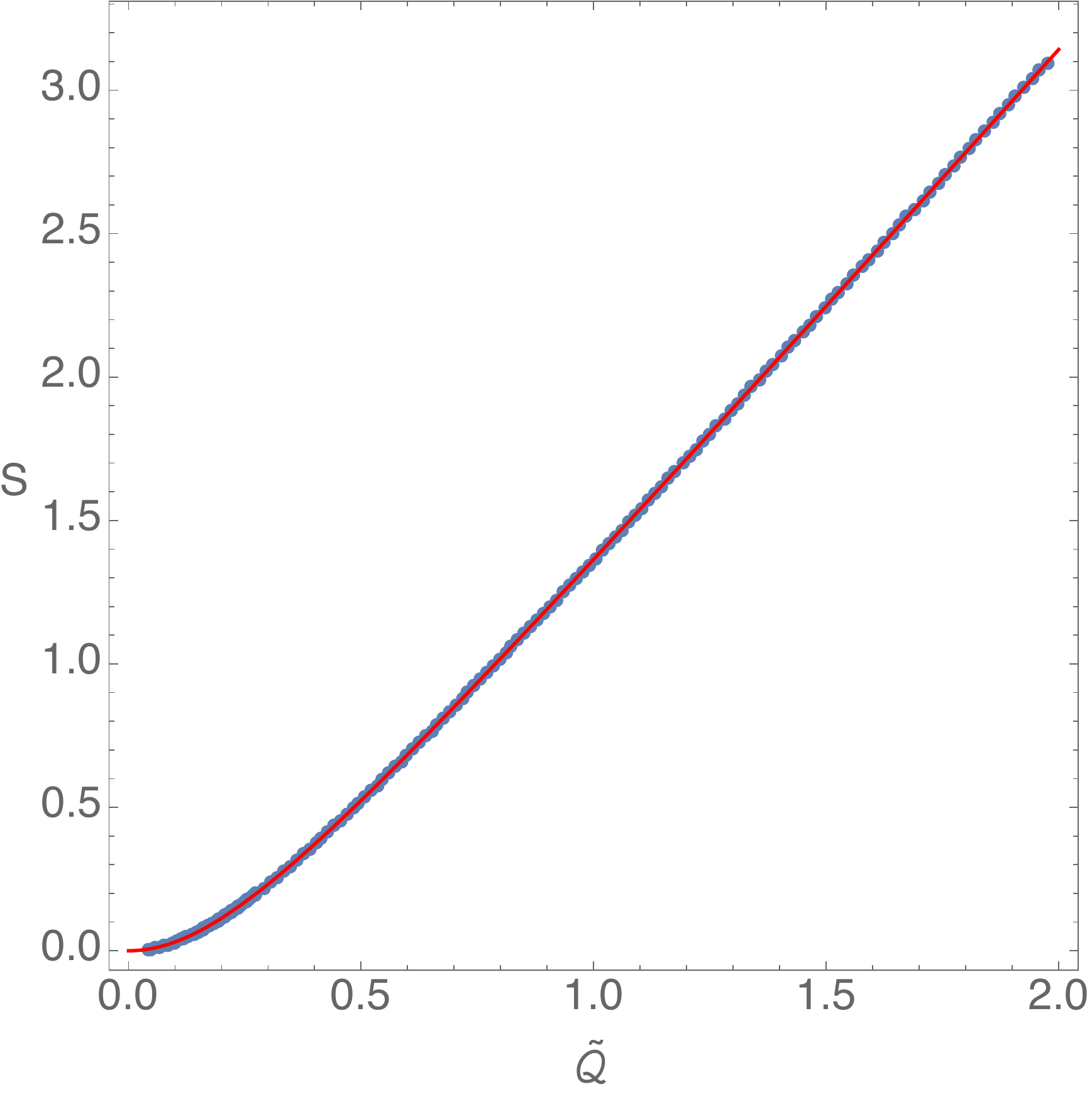}
\end{center}
\caption{\label{fig:dis}Entropy of the hovering black hole as a function of the total charge enclosed in a hovering horizon. The blue dots are our numerical data, and the dashed red curve is the analytic prediction (\ref{eq:entropycharge}). This particular example was generated using profile $I_1$ in Eq.~\ref{eq:profiles}.}
\end{figure}

First, let us examine the near horizon geometry of the hovering black holes.  We find that they match that of spherically symmetric extremal Reissner-N\"{o}rdstrom black holes in AdS.  If our solutions are smooth, this is required by a classification theorem of near-horizon geometries \cite{Kunduri:2013gce}.  For instance, the entropy of these black holes as a function of its total charge matches that of Reissner-N\"{o}rdstrom:
\begin{equation}
\mathcal S = \frac{L^2}{6} \pi  \left(\sqrt{12 \mathcal Q^2+1}-1\right)\,.
\label{eq:entropycharge}
\end{equation}
In Fig.~\ref{fig:dis} we plot the entropy of the numerical hovering black hole (blue dots) as a function of the charge contained inside the horizon. The dashed red line is the analytic prediction (\ref{eq:entropycharge}). The agreement is excellent.  We note that we have not required this as an input to our numerics, so this is a reassuring numerical check.

It is amusing to ask where the flux from the charged black hole goes in the case where $\mu(r)$ is an irrelevant deformation. We have seen in Sec. \ref{sec:falloffs}\ that the total charge at infinity is zero, and the IR geometry consists of a Poincar\'{e} horizon which has no charge.  So in Fefferman-Graham coordinates, the flux must all escape through the ``sides" (in our coordinates, this is the boundary at large distance from the axis of symmetry). 

Note that there is a region in parameter space where the hovering black hole solutions coexist with the solutions without the black hole, namely for $a\in(a_{\star},a_{\max})$.  It is therefore natural to ask which of these phases will dominate in a particular ensemble. In the micro-canonical ensemble, it is clear that the solutions with hovering black holes will dominate because they have nonvanishing entropy. Furthermore, the phase transition at $a=a_{\star}$ is second order because there, the black holes have zero size. This in turn indicates that, at any non-zero but small temperature, the hovering solutions dominate all ensembles, because they are guaranteed to dominate the micro-canonical ensemble and the transition is second order \cite{secondorder}.  We also point out that this is a quantum phase transition since our solutions are at zero temperature. The transition is localized in space, as it involves a total entropy which does not scale with volume. Thus (as is usual in holography) the thermodynamic limit required for a phase transition is provided not by infinite volume, but by the fact that we are working at large $N$. 

The size of the hovering black holes increases monotonically as we increase the amplitude $a$ and can become quite large. If we define a horizon radius by
 $r_+ = \sqrt{\mathcal S/\pi}$, then we find that it can be larger than the AdS length scale $L$.  In fact, we have reached sizes as large as $r_+ \sim 3\,L$. 
 
\begin{figure}[th]
\begin{center}
\includegraphics[width=\textwidth]{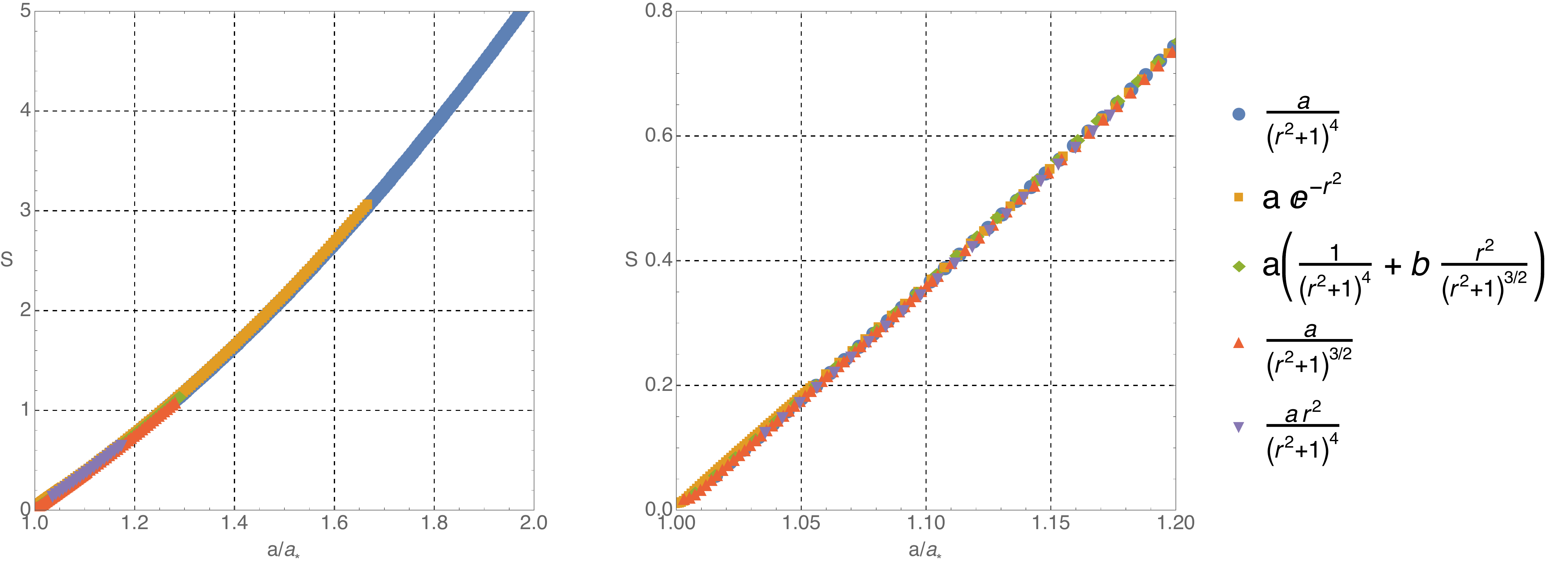}
\end{center}
\caption{\label{fig:entropy}Entropy of the hovering black hole as a function of $a/a_{\star}$ for several boundary profiles. The left panel shows all of our data points, while the right panel shows a zoom of our data close to $a/a_{\star}\sim1$. The different symbols used indicate the various profiles we have considered, which we label on the right.  Here we have chosen $b=0.075$.}
\end{figure}

 As mentioned in the introduction, the way in which the hovering black holes grow with $a/a_{\star}$ appears to exhibit a remarkable universality. It seems completely independent of the profile of the chemical potential. The plot on the left of Fig.~\ref{fig:entropy} shows the entropy of the hovering black hole as a function of $a/a_{\star}$ for five different boundary profiles. The fact that the curves agree is surprising and not understood. The plot on the right is a close-up of the data for small black holes, and shows a clear linear dependence.  The analogous curve for extremal RN black holes in global AdS is given by \eqref{eq:entropy}.  This also has a linear scaling for small black holes, but the slope is different and \eqref{eq:entropy} is nowhere a good approximation to  Fig.~\ref{fig:entropy}.   This is not surprising since the boundary geometry is $\mathbb R\times S^2$ rather than Minkowski space, and the chemical potential is constant.

\section{Discussion \& Outlook}

We have constructed the holographic dual of a localised electrically charged  defect in a strongly coupled conformal field theory.    When the strength of the defect is large enough, the dual gravitational description contains a charged black hole hovering above the IR horizon. This hovering black hole can have unusual consequences in the dual field theory. One obvious consequence is that there must be  a large number of degenerate states localised near the defect.  Since the hovering black hole is in the middle of the bulk (not in the UV or IR part of the spacetime) it will have its biggest effect at intermediate or ``mid-infrared" energies. It is as if the degenerate states all have a similar characteristic size. Imagine sending a signal in the dual field theory toward the defect. At high energies the signal will pass right through, corresponding to a perturbation of  the bulk that passes above the hovering black hole. At low energies, the signal will also pass by largely unchanged, corresponding to a perturbation of  the bulk that passes below the hovering black hole. For a suitable range of intermediate energies, the perturbation hits the black hole and is largely absorbed.  The resulting black hole is now slightly non-extremal and will Hawking evaporate (at finite $N$) back to extremality. In the dual field theory, this corresponds to  the signal ``rapidly thermalising" with the degenerate states around the defect, and then radiating away the excess heat. One should see a dramatic signature of this as one scans in energy. It would be interesting to investigate this phenomenon further.

Perhaps the most important open question is to understand the universal behaviour we see for the entropy as a function of the strength of the defect  discussed above.  It would help enormously to have an analytic solution of this type. Analytic solutions are indeed known in slightly different contexts. There is a charged C-metric that describes a charged black hole  uniformly accelerating in a spacetime with no cosmological constant \cite{Kinnersley:1970zw}. The source of the acceleration is a conical singularity along the axis which acts like a cosmic string pulling the black hole. One can remove the conical singularity by adding a background electric flux tube which can source the acceleration \cite{Ernst:1976}.  We need a solution of this type with nonzero $\Lambda$. There is indeed a generalisation of the charged C-metric to include  $\Lambda \ne 0$ \cite{Plebanski:1976gy}, but it is apparently not known how to add the background electric flux tube analytically.

This universal behaviour is reminiscent of Choptuik scaling \cite{Choptuik:1992jv}, which concerns small black holes formed from, e.g., the collapse of spherical scalar fields. It was found that near the threshold for black hole formation, the size of the black hole scaled with the initial amplitude of the scalar field in a universal way that was independent of the initial radial profile. The universality we find for hovering black holes has a similar character, but there are two important differences. First, we are considering static, zero temperature ground states, not dynamical collapse. Second, and more importantly,  our universality extends to large black holes, not just very small ones.

We have not discussed the case of a relevant deformation, where $\mu(r)$ falls off more slowly than $1/r$. We have found finite temperature solutions without a black hole, but the IR appears to become singular as $T \to 0$.  At finite temperature, there are certain profiles that permit hovering black holes, but these static orbits do not persist down to zero temperature.  

There are several new directions that are worth exploring.  Rather than a single isolated defect, one could consider an array of defects. When their amplitude is small,  the IR will be described by an extremal charged horizon. (In this case there is no complication associated with the fall-off of the chemical potential as we saw for a single isolated defect, though these horizons might be highly non-analytic which would pose a technical challenge.) As one increases the strength of the defects, one expects spherical charged black holes to develop, hovering above the horizon. Eventually, one can imagine that all the charge will be contained in the localized black holes and none will remain on the IR horizon. This may be dual to localization in the presence of strong defects.

Besides an array of defects, one can also attempt to find axisymmetric solutions with multiple hovering black holes at various distances into the bulk.  A study of charged geodesics in our \emph{hovering} solutions suggests that our profiles do not permit additional hovering black holes, but this might still be accomplished by introducing a profile with additional length scales.  

We have focussed mostly on zero temperature solutions, but $T>0$ solutions with nonextremal hovering black holes should exist as well. If one now includes a charged scalar field in the theory, then  at low temperature, the scalar field will condense around the charged black holes. This is just a localized version of the instability resulting in the holographic superconductor \cite{Hartnoll:2008vx}. Nuggets of the charged condensate $\langle {\cal O}\rangle$ will form in the dual theory. Even in the phase without black holes, if the electric field in the bulk is strong enough along the axis $r=0$, one might expect there to be an instability to forming a nonzero charged scalar field there. 

From a purely gravitational standpoint, the existence of hovering black holes is an interesting new type of black hole solution. One might wonder if analogous solutions exist for neutral black holes. One could imagine constructing such a solution by replacing the flat Minkowski metric with a  suitable inhomogeneous boundary metric (or taking gravity coupled to a neutral scalar field with an inhomogeneous source). The resulting bulk solution might contain a static geodesic above the Poincar\'{e} horizon. If so, one could add a small Schwarzschild black hole at the location of the geodesic and it should remain static.
 
\vskip 1cm
\centerline{\bf Acknowledgements}
\vskip .5 cm
We thank Mike Blake, Jan de Boer, Aristomenis Donos, Sean Hartnoll and David Tong for discussions. G. H. was supported in part by NSF grant PHY12-05500. The research leading to these results has received funding from the European Research Council under the European Community's Seventh Framework Programme (FP7/2007-2013) / ERC grant agreement no. [247252]. N.I. is supported by the D-ITP consortium, a program of the Netherlands Organisation for Scientific Research (NWO) that is funded by the Dutch Ministry of Education, Culture and Science (OCW). B.W. is supported by European Research Council grant no. ERC-2011-StG 279363-HiDGR. N.T.
\vskip 1cm
{\bf Note added in proof:} in Section \ref{sec:geodesics} it was argued that hovering black holes can form only when the minimum $\sV_{min}$ of the effective potential felt by a test particle drops below zero. This is actually only the case when there is vanishing potential difference $\Delta \mu$ between the hovering black hole horizon and the Poincare horizon. We have recently checked that hovering black holes also exist with nonzero potential difference, in which case the analysis in Section \ref{sec:geodesics} does not immediately apply.


\appendix

\section{\label{app:1}Perturbative results for small amplitude defects}  In this appendix we give the details of our perturbative calculation for small amplitudes about $AdS_4$.  While the specifics of this calculation depend upon the profile, the profiles that decay faster than $1/r^2$ at large $r$ are very similar. For simplicity, we will only detail the perturbative calculation for the profile $I_1$ to second order in $a$. 

We begin by writing our line element and gauge field in Fefferman-Graham coordinates:
\begin{subequations}
\begin{equation}
\dd s^2 = \frac{L^2}{z^2}\left[-G(r,z)\dd t^2+B(r,z)\dd r^2+C(r,z)r^2 \dd \phi^2+dz^2\right]\,
\end{equation}
and
\begin{equation}
A = A_t(r,z)\,\dd t\,,
\end{equation}
\end{subequations}
where $G$, $B$, $C$ and $A_t$ are functions of $r$ and $z$ to be determined in what follows. We then expand our functions in powers of $a$:
\begin{align}
&G = 1+\sum_{i=1}^{+\infty}a^{2i} G^{(2i)}(r,z)\nonumber\\
&B = 1+\sum_{i=1}^{+\infty}a^{2i} B^{(2i)}(r,z)\nonumber\\
&C = 1+\sum_{i=1}^{+\infty}a^{2i} C^{(2i)}(r,z)\nonumber\\
&A_t = \sum_{i=0}^{+\infty}a^{2i+1} A_t^{(2i+1)}(r,z)\,.\nonumber
\end{align}
Notice that to first order in $a$, only $A_t$ is non-trivial, so we can solve this order using just the Maxwell equations.

The Maxwell equations yields the following linear equation for $A_t^{(1)}$:
\begin{equation}
\frac{\partial^2 A_t^{(1)}}{\partial z^2}+\frac{\partial^2 A_t^{(1)}}{\partial r^2}+\frac{1}{r}\frac{\partial A_t^{(1)}}{\partial r}=0\,.
\label{eq:linear}
\end{equation}
Our objective is to solve Eq.~(\ref{eq:linear}) subject to our boundary profile:
\begin{equation}
a A_t^{(1)}(r,0) =  \mu_{I_1}(r) = \frac{a}{\left(\frac{r^2}{\ell^2}+1\right)^{3/2}}\,.
\label{eq:proapp}
\end{equation}
We start by assuming a separable solution of the form
\begin{equation}
A_t^{(1)}(r,z) = R(r)Z(z)\,.
\end{equation}
Which, upon imposing bulk regularity, leads to
\begin{equation}
Z(z) = e^{- k\, z}\,\quad \text{and}\quad R(r) = J_0(k r)\,,
\end{equation}
where $k$ is a separation constant, and $J_0$ is the zeroth order Bessel function of the first kind. The general smooth solution of Eq.~(\ref{eq:linear}) can then be written as
\begin{equation}
A_t^{(1)}(r,z) = \int_{0}^{+\infty}\dd k\,k f(k)\,J_0(k r)\,e^{- k\,z}\,
\label{eq:linearsol}
\end{equation}
where $f$ is a function determined by boundary conditions. Using the fact that Bessel functions form a basis on the semi-infinite line, we can further observe the following
\begin{equation}
\int_{0}^{+\infty} \dd r\,r\,J_0(k r)\,J_0(k^{\prime}r) = \frac{\delta(k-k^{\prime})}{k}\quad\text{and}\quad \int_{0}^{+\infty} \dd k\,k\,J_0(k r)\,J_0(kr^{\prime}) = \frac{\delta(r-r^{\prime})}{r}\,,
\end{equation}
for $k,\,k^{\prime},\,r,\,r^{\prime}\in\mathbb{R}^+$. Which in turn implies that:
\begin{equation}
f(k) = \int_{0}^{+\infty}\dd r\,r\,J_0(k r)A_t^{(1)}(r,0)\,.
\end{equation}
This together with Eq. \eqref{eq:linearsol} gives us our solution to first order in $a$.  

For the simple profile (\ref{eq:proapp}) one finds
\begin{equation}
f(k) = \ell^2 e^{-\ell\,k}\,,
\end{equation}
which, together with Eq.~(\ref{eq:linearsol}), gives
\begin{equation}
a A_t^{(1)}(r,z) = \frac{a\,\ell ^2 (z+\ell )}{\left[r^2+(z+\ell )^2\right]^{3/2}}\,.
\label{eq:sollinear1}
\end{equation}
Expanding this equation in $z$ gives us the charge density to first order in $a$:
\begin{equation}
\rho(r) = -\frac{a \ell ^2 \left(r^2-2 \ell ^2\right)}{4 \pi  \left(r^2+\ell ^2\right)^{5/2}}\,,
\end{equation}
which at large $r$ decays like $-(a \ell ^2)/(4 \pi  r^3)$, as expected. A simple integration tells us that the total charge $Q$ is zero. 

We can construct new linear solutions for different profiles by exploiting the fact that Eq.~(\ref{eq:linear}) is linear.  In particular, we can find what happens for a generic Maxwell field whose boundary behaviour is given by
\begin{equation}
a A_t^{(1)}(r,0) = \frac{a}{\left(\frac{r^2}{\ell^2}+1\right)^{\frac{3}{2}+n}} = \frac{\sqrt{\pi } (-1)^n 2^{-n-1} \ell ^{2 n+3}}{\Gamma \left(n+\frac{3}{2}\right)}\mathcal{D}^n\left(\frac{1}{\ell^3}\mu_{I_{1}}\right)\,\quad\text{for}\quad n\in \mathbb{N}\,,
\end{equation}
where $\mathcal{D}(f)\equiv \ell^{-1}\partial f/\partial \ell$, which gives
\begin{equation}
a A_t^{(1)}(r,z) = \frac{\sqrt{\pi } (-1)^n 2^{-n-1} \ell ^{2 n+3}}{\Gamma \left(n+\frac{3}{2}\right)}\mathcal{D}^n\left(\frac{1}{\ell^3}\frac{a\,\ell ^2 (z+\ell )}{\left[r^2+(z+\ell )^2\right]^{3/2}}\right)\,\quad\text{for}\quad n\in \mathbb{N}\,.
\end{equation}

We now proceed to second order. The calculation gets substantially more involved, so we continue to discuss our profile $I_1$ and henceforth keep $n=0$. At this order, the Maxwell equations are already satisfied, and we need to solve the Einstein equations.  We first define the function $q =G^{(2)}-C^{(2)}$, which we will use to replace the function $G^{(2)}$. From the $zz$ component of Eq.~(\ref{eq:einsteinmaxwell}) we find
\begin{multline}
B^{(2)} =\frac{z^2 C^\prime_1(r)}{2\,r}+C_2(r)+\frac{1}{64} \Bigg\{\frac{9 \ell ^5 (z^2-\ell^2)\arctan\left(\frac{z+\ell }{r}\right)}{r^5}+\frac{9 \ell ^5 (z+\ell )}{r^4}-\frac{9 \ell ^6}{r^4}-\frac{21 \ell ^5 \arctan\left(\frac{z+\ell }{r}\right)}{r^3}+\\
\frac{48 r^2 \ell ^4 \left[r^2+\ell  (2 z+\ell )\right]}{\left[r^2+(z+\ell )^2\right]^3}-\frac{2 \ell ^4 \left[3 \ell  (4 z+3 \ell )-32 r^2\right]}{r^2 \left[r^2+(z+\ell)^2\right]}-\frac{4 \ell ^4 \left[28 r^2+\ell  (23 z+16 \ell )\right]}{\left[r^2+(z+\ell )^2\right]^2}\Bigg\}-q(r,z)\,,
\label{eq:b2eq}
\end{multline}
where $C_1$, and $C_2$ are arbitrary integration functions. It turns out that the $rz$ component of Eq.~(\ref{eq:einsteinmaxwell}) can also be integrated, to express $C^{(2)}$ in terms of $q$:
\begin{multline}
C^{(2)}=-r \frac{\partial q}{\partial r}-q-\frac{1}{32 r^5}\Bigg\{16 r^4 z^2 C_1'(r)+\frac{3}{2} \ell ^5 \left(11 r^2+3 \ell ^2\right) \arctan\left(\frac{r}{z+\ell }\right)-
\\
\frac{144 r^9 \ell ^4 \left[r^2+\ell  (2 z+\ell )\right]}{\left[r^2+(z+\ell)^2\right]^4}+\frac{8 r^7 \ell ^4 \left[43 r^2+\ell  (44 z+25 \ell )\right]}{\left[r^2+(z+\ell )^2\right]^3}-\frac{2 r^5 \ell ^4 \left[124 r^2+\ell  (25 z+18 \ell )\right]}{\left[r^2+(z+\ell )^2\right]^2}+
\\
\frac{r^3 \ell ^4\left[32 r^2-3 \ell  (6 z+5 \ell )\right]}{r^2+(z+\ell )^2}+\frac{9}{2} z^2 \ell ^5 \arctan\left(\frac{z+\ell }{r}\right)+\frac{9}{2} r z \ell ^5\Bigg\}+C_3(r)\,,
\label{eq:c2eq}
\end{multline}
where $C_3$ is another integration function.  The trace of Eq.~(\ref{eq:einsteinmaxwell}) expresses $C_2(r)$ as a function of $C_1(r)$ and $C_3(r)$:
\begin{equation}
C_2(r) = -2 C_1(r)+C_3(r)+\kappa_0+\frac{9 \pi  \ell ^7}{128 r^5}+\frac{33 \pi  \ell ^5}{128 r^3}\,,
\end{equation}
where $\kappa_0$ is an integration constant. The $rr$ component of Eq.~(\ref{eq:einsteinmaxwell}) gives the only ``dynamical'' equation to solve for $q$, which takes the following form:
\begin{multline}
\frac{\partial^2q}{\partial z^2}-\frac{2}{z}\frac{\partial q}{\partial z}+\frac{\partial^2q}{\partial r^2}+\frac{3}{r}\frac{\partial q}{\partial r}
\\
-\frac{9 \ell ^5}{64 r^7}\left(11 r^2-5 z^2+5 \ell ^2\right) \arctan\left(\frac{z+\ell}{r}\right)-\frac{2 r^2 C_3^{\prime}(r)+z^2 \left[r C_1^{\prime\prime}(r)-C_1^{\prime}(r)\right]}{2 r^3}+
\\
\frac{\ell ^4}{128 r^7 \left[r^2+(z+\ell )^2\right]^4}\Bigg[99 \pi  r^{10} \ell-128 r^{11} -2 r^9 \left(320 z^2+413 z \ell +185 \ell ^2\right)+
\\
9 \pi  r^8 \ell  \left(44 z^2+88 z \ell +49 \ell ^2\right)+4 r^7 (z+\ell )^2 \left(304 z^2+63 z \ell -188 \ell ^2\right)+
\\
18 \pi r^6 \ell  (z+\ell )^2 \left(33 z^2+66 z \ell +43 \ell ^2\right)-12 r^5 \ell  (z+\ell )^4 (24 z+61 \ell )+
\\
18 \pi  r^4 \ell  (z+\ell )^4 \left(22 z^2+44 z \ell +37 \ell ^2\right)+12 r^3 \ell  (11 z-20 \ell ) (z+\ell )^6\\
+9 \pi r^2 \ell  (z+\ell )^6 \left(11 z^2+22 z \ell +31 \ell ^2\right)+18 r \ell  (5 z-\ell ) (z+\ell )^8+45 \pi  \ell ^3 (z+\ell )^8\Bigg]\,.
\label{eq:qeq}
\end{multline}
The remaining component of the Einstein equations is automatically satisfied if we solve \eqref{eq:b2eq}, \eqref{eq:c2eq}, and \eqref{eq:qeq}.

Let us now simplify the complicated equation \eqref{eq:qeq} by a number of redefinitions.  We first parametrise $C_1$ and $C_3$ as:
\begin{align}
C_1(r)\equiv r \left(3 \lambda _0^{\prime}+r \lambda _0^{\prime\prime}\right)\nonumber
\\
C_3(r) \equiv r \lambda _1^{\prime}+2 \lambda_1\,\nonumber
\end{align}
where $\lambda_0$ and $\lambda_1$ are arbitrary functions of $r$. If we redefine $q$ as
\begin{equation}
q = \lambda_0+\lambda_1+\frac{z^2\,C_1}{2r^2}+\frac{1}{r}\frac{\partial}{\partial_r}\left(\tilde{q}-z\frac{\partial \tilde{q}}{\partial z}\right)
\end{equation}
and then redefine $\tilde q$ as 
\begin{multline}
\tilde{q} = \widehat{q}+\frac{\pi  \ell ^6 (z+\ell )}{128 r^3}+\frac{5 \ell ^5 (z+\ell )}{256 r^2}+\frac{1}{2} \ell ^4 \log \left(\sqrt{r^2+(z+\ell )^2}\right)+
\\
\frac{1}{r^2+(z+\ell )^2}\left[\frac{1}{32} \ell ^4 \left(5 r^2+3 \ell ^2\right)+\frac{3 z \ell^5}{32}+\frac{1}{16} \ell ^4 (z+\ell )^2\right]-
\\
\left[\frac{z^2 \ell ^5}{64 r^3}+\frac{z \left(13 r^2 \ell ^4+\ell ^6\right)}{32 r^3}+\frac{31 r^2 \ell ^5+\ell ^7}{64 r^3}\right]\arctan\left(\frac{z+\ell}{r}\right)+\frac{1}{r}\left(\frac{13}{64} \pi  z \ell ^4+\frac{33 \pi  \ell ^5}{128}\right)\,,
\end{multline}
we find that \eqref{eq:qeq} reduces to
\begin{equation}
\frac{\partial^2\widehat{q}}{\partial z^2}+\frac{\partial^2\widehat{q}}{\partial r^2}+\frac{1}{r}\frac{\partial \widehat{q}}{\partial r}=0\,,
\end{equation}
which is actually the same as Eq.~\eqref{eq:linear}.  It therefore has the solution
\begin{equation}
q(r,z) = \int_{0}^{+\infty}\dd k\,k g(k)\,J_0(k r)\,e^{- k\,z}\,,
\end{equation}
for some integration function $g$.  We have thus solved the equations to second order in $a$, aside from the undetermined functions $\lambda_0$, $\lambda_1$, $g$, and the constant $\kappa_0$. 

The remainder of the calculation is completed by imposing boundary conditions.  Apart from regularity everywhere in the bulk, we must choose $\widehat{q}(r,0)$ carefully, to ensure that our boundary metric is conformally flat. This condition translates into finding $\widehat{q}$ such that
\begin{equation}
G^{(2)}(r,0) = B^{(2)}(r,0)= C^{(2)}(r,0)\,,
\end{equation}
which allows us to express $\lambda_1$ and $\widehat{q}(r,0)$ in terms of elementary functions. We will omit from the text what $\lambda_1$ turns out to be, but for reference we present here $\widehat{q}$
\begin{equation}
\widehat{q}(r,0) = \frac{3}{32} \ell ^4 \text{Li}_2\left(-\frac{r^2}{\ell ^2}\right)+\frac{5}{64} \ell ^4 \log \left(\frac{r^2}{\ell ^2}+1\right)\,
\end{equation}
where $\text{Li}_2$ is the dilogarithm function. Note that the expression for $\widehat{q}$ everywhere in the bulk will involve an integral form similar to Eq.~(\ref{eq:linearsol}), which we omit here. After fixing all integration variables by demanding regularity we can finally read off the stress energy tensor. Since the stress energy tensor must obey the Ward identities and is traceless, there is only one independent component. Here we choose to present $T^{t}_{\phantom{t}t}$ for which we obtain:
\begin{equation}
T^{t}_{\phantom{t}t} = \frac{\mathcal{A}^2}{512\pi\,\ell}\mathfrak{f}\left(\frac{r}{\ell}\right),
\end{equation}
with
\begin{multline}
\mathfrak{f}(x) = \frac{1}{\left(1+x^2\right)^5}\Large[12 \left(1+x^2\right) \left(9 x^4-51 x^2+20\right) K\left(-x^2\right)\\
+128 \left(5 x^4-8 x^2+2\right)-3 \left(111 x^4-410 x^2+119\right) E\left(-x^2\right)\Large]\,,
\end{multline}
where $K(y)$ and $E(y)$ are the complete Elliptic integral of the first kind and complete Elliptic integral, respectively.

\bibliography{refs}{}

\providecommand{\href}[2]{#2}\begingroup\raggedright\begin{thebibliography}{10}

\bibitem{Harrison:2011fs}
S.~Harrison, S.~Kachru, and G.~Torroba, {\it {A maximally supersymmetric Kondo
  model}},  {\em Class.Quant.Grav.} {\bf 29} (2012) 194005,
  [\href{http://xxx.lanl.gov/abs/1110.5325}{{\tt arXiv:1110.5325}}].

\bibitem{Zeng:2014dra}
H.~B. Zeng and H.-Q. Zhang, {\it {Zeroth Order Phase Transition in a
  Holographic Superconductor with Single Impurity}},
  \href{http://xxx.lanl.gov/abs/1411.3955}{{\tt arXiv:1411.3955}}.

\bibitem{Erdmenger:2013dpa}
J.~Erdmenger, C.~Hoyos, A.~Obannon, and J.~Wu, {\it {A Holographic Model of the
  Kondo Effect}},  {\em JHEP} {\bf 1312} (2013) 086,
  [\href{http://xxx.lanl.gov/abs/1310.3271}{{\tt arXiv:1310.3271}}].

\bibitem{Blake:2014lva}
M.~Blake, A.~Donos, and D.~Tong, {\it {Holographic Charge Oscillations}},
  \href{http://xxx.lanl.gov/abs/1412.2003}{{\tt arXiv:1412.2003}}.

\bibitem{Hickling:2014dra}
A.~Hickling, J.~Lucietti, and T.~Wiseman, {\it {Null infinity and extremal
  horizons in AdS-CFT}},  \href{http://xxx.lanl.gov/abs/1408.3417}{{\tt
  arXiv:1408.3417}}.

\bibitem{Jensen:2011su}
K.~Jensen, S.~Kachru, A.~Karch, J.~Polchinski, and E.~Silverstein, {\it
  {Towards a holographic marginal Fermi liquid}},  {\em Phys.Rev.} {\bf D84}
  (2011) 126002, [\href{http://xxx.lanl.gov/abs/1105.1772}{{\tt
  arXiv:1105.1772}}].

\bibitem{metlitski2008valence}
M.~A. Metlitski and S.~Sachdev, {\it Valence bond solid order near impurities
  in two-dimensional quantum antiferromagnets},  {\em Physical Review B} {\bf
  77} (2008), no.~5 054411, [\href{http://xxx.lanl.gov/abs/0710.0626}{{\tt
  arXiv:0710.0626}}].

\bibitem{Dias:2013bwa}
O.~J. Dias, G.~T. Horowitz, N.~Iqbal, and J.~E. Santos, {\it {Vortices in
  holographic superfluids and superconductors as conformal defects}},  {\em
  JHEP} {\bf 1404} (2014) 096, [\href{http://xxx.lanl.gov/abs/1311.3673}{{\tt
  arXiv:1311.3673}}].

\bibitem{Cai:2004pz}
R.-G. Cai and A.~Wang, {\it {Thermodynamics and stability of hyperbolic charged
  black holes}},  {\em Phys.Rev.} {\bf D70} (2004) 064013,
  [\href{http://xxx.lanl.gov/abs/hep-th/0406057}{{\tt hep-th/0406057}}].

\bibitem{Kunduri:2013gce}
H.~K. Kunduri and J.~Lucietti, {\it {Classification of near-horizon geometries
  of extremal black holes}},  {\em Living Rev.Rel.} {\bf 16} (2013) 8,
  [\href{http://xxx.lanl.gov/abs/1306.2517}{{\tt arXiv:1306.2517}}].

\bibitem{Ryu:2006bv}
S.~Ryu and T.~Takayanagi, {\it {Holographic derivation of entanglement entropy
  from AdS/CFT}},  {\em Phys.Rev.Lett.} {\bf 96} (2006) 181602,
  [\href{http://xxx.lanl.gov/abs/hep-th/0603001}{{\tt hep-th/0603001}}].

\bibitem{PhysRevLett.67.161}
I.~Affleck and A.~W.~W. Ludwig, {\it Universal noninteger ``ground-state
  degeneracy'' in critical quantum systems},  {\em Phys. Rev. Lett.} {\bf 67}
  (Jul, 1991) 161--164.

\bibitem{Jensen:2013lxa}
K.~Jensen and A.~O'Bannon, {\it {Holography, Entanglement Entropy, and
  Conformal Field Theories with Boundaries or Defects}},
  \href{http://xxx.lanl.gov/abs/1309.4523}{{\tt arXiv:1309.4523}}.

\bibitem{Anninos:2013mfa}
D.~Anninos, T.~Anous, F.~Denef, and L.~Peeters, {\it {Holographic
  Vitrification}},  \href{http://xxx.lanl.gov/abs/1309.0146}{{\tt
  arXiv:1309.0146}}.

\bibitem{Headrick:2009pv}
M.~Headrick, S.~Kitchen, and T.~Wiseman, {\it {A New approach to static
  numerical relativity, and its application to Kaluza-Klein black holes}},
  {\em Class.Quant.Grav.} {\bf 27} (2010) 035002,
  [\href{http://xxx.lanl.gov/abs/0905.1822}{{\tt arXiv:0905.1822}}].

\bibitem{Figueras:2011va}
P.~Figueras, J.~Lucietti, and T.~Wiseman, {\it {Ricci solitons, Ricci flow, and
  strongly coupled CFT in the Schwarzschild Unruh or Boulware vacua}},  {\em
  Class.Quant.Grav.} {\bf 28} (2011) 215018,
  [\href{http://xxx.lanl.gov/abs/1104.4489}{{\tt arXiv:1104.4489}}].

\bibitem{Balasubramanian:1999re}
V.~Balasubramanian and P.~Kraus, {\it {A Stress tensor for Anti-de Sitter
  gravity}},  {\em Commun.Math.Phys.} {\bf 208} (1999) 413--428,
  [\href{http://xxx.lanl.gov/abs/hep-th/9902121}{{\tt hep-th/9902121}}].

\bibitem{deHaro:2000xn}
S.~de~Haro, S.~N. Solodukhin, and K.~Skenderis, {\it {Holographic
  reconstruction of space-time and renormalization in the AdS / CFT
  correspondence}},  {\em Commun.Math.Phys.} {\bf 217} (2001) 595--622,
  [\href{http://xxx.lanl.gov/abs/hep-th/0002230}{{\tt hep-th/0002230}}].

\bibitem{secondorder}
F.~Bouchet and J.~Barr\'e, {\it Classification of phase transitions and
  ensemble inequivalence, in systems with long range interactions},  {\em
  Journal of Statistical Physics} {\bf 118} (2005), no.~5-6 1073--1105.

\bibitem{Kinnersley:1970zw}
W.~Kinnersley and M.~Walker, {\it {Uniformly accelerating charged mass in
  general relativity}},  {\em Phys.Rev.} {\bf D2} (1970) 1359--1370.

\bibitem{Ernst:1976}
J.~Ernst, {\it {Removal of the nodal singularity of the C-metric}},  {\em J.
  Math Phys.} {\bf 17} (1976) 515.

\bibitem{Plebanski:1976gy}
J.~Plebanski and M.~Demianski, {\it {Rotating, charged, and uniformly
  accelerating mass in general relativity}},  {\em Annals Phys.} {\bf 98}
  (1976) 98--127.

\bibitem{Choptuik:1992jv}
M.~W. Choptuik, {\it {Universality and scaling in gravitational collapse of a
  massless scalar field}},  {\em Phys.Rev.Lett.} {\bf 70} (1993) 9--12.

\bibitem{Hartnoll:2008vx}
S.~A. Hartnoll, C.~P. Herzog, and G.~T. Horowitz, {\it {Building a Holographic
  Superconductor}},  {\em Phys.Rev.Lett.} {\bf 101} (2008) 031601,
  [\href{http://xxx.lanl.gov/abs/0803.3295}{{\tt arXiv:0803.3295}}].

\end{thebibliography}\endgroup
\bibliographystyle{JHEP}

\end{document}